\documentclass[aps,twocolumn,superscriptaddress,floatfix,amsmath,a4paper,showkeys]{revtex4-1}
\usepackage{graphicx}
\usepackage{ulem}
\usepackage[utf8]{inputenc}
\pdfoutput=1
\begin{document}

\title{Anomalous diffusion: A basic mechanism for the evolution of inhomogeneous  systems}

\author{Fernando A. Oliveira}
\email{fao@fis.unb.br}
\affiliation{Institute of Physics, Universidade de Bras\'{\i}lia, Bras\'{\i}lia, Brazil.}
\author{Rogelma M. S. Ferreira}
\affiliation{Centro de Ci\^encias Exatas e Tecnol\'ogicas, Universidade Federal do Rec\^oncavo da Bahia, Cruz das Almas, Bahia, Brazil}
\author{Luciano C. Lapas}
\affiliation{Interdisciplinary Center for Natural Sciences, UNILA, P.O. Box 2067, 85867-970 Foz do Igua\c{c}u, Brazil.}
\author{Mendeli H. Vainstein}
\affiliation{Institute of Physics, Universidade Federal do Rio Grande do Sul, Porto Alegre, Brazil.}
\affiliation{Physics of Livings Systems, Department of Physics, Massachusetts Institute of Technology, Cambridge, MA, United States of America}

\begin{abstract}
 In this article we review classical and recent results in anomalous diffusion and provide mechanisms useful for the study of the fundamentals of certain processes, mainly in condensed matter physics, chemistry and biology. Emphasis will be given to some methods applied in the analysis and characterization of diffusive regimes through the memory function,  the mixing condition (or irreversibility), and ergodicity. Those methods can be used in the study of small-scale systems, ranging in size from single-molecule to particle clusters and including among others polymers, proteins, ion channels and biological cells, whose diffusive properties have received much attention lately.
\vspace{0.2cm}

\end{abstract}


\maketitle

\onecolumngrid
\begin{quote}
\flushright{``In water, this flowing matter, there is nothing permanent" \\ Jan Heweliusz}
\end{quote} 
\vspace{0.5cm}
\twocolumngrid

\section{Introduction}

\subsection{General concepts}
Diffusion is a basic transport process involved in the evolution of many non-equilibrium systems towards equilibrium \cite{Vainstein06,Shlesinger93,Metzler99,Metzler00,Morgado02,Metzler04,Sancho04,Costa03,Lapas08, Weron10,Thiel13, Mckinley18,Flekkoy17}. By diffusion, particles (or molecules) spread from regions of high concentration to those of low concentration leading, via a gradual mixing, to a situation in which they become evenly dispersed. Diffusion is of fundamental importance in many disciplines; for example, in growth phenomena, the Edwards-Wilkinson equation is given by a diffusion equation plus a noise~\cite{Barabasi95}. In cell biology it constitutes  a main form of transport for amino acids and other nutrients within cells~\cite{Murray2002}.

For more than two hundred years \cite{Vainstein06}, diffusion  has been a widely studied phenomena in natural science due to it's large number of applications. Initially, the experiments carried out by Robert Brown~\cite{Brown28,Brown28a}  called the attention to the random trajectories of small particles of polen and also of inorganic matter. This irregular motion, later  named Brownian motion, can be modeled by a random walk in which the mean square displacement is given by Einstein’s relation \cite{Vainstein06}
\begin{equation}
\label{eq.sqr_disp}
 \langle (\Delta r )^2 \rangle = 2dDt,
\end{equation}
 where $\Delta r$ is the displacement of the Brownian particle in a given time interval $t$, $d$ is the spatial  dimension, and $D$ the diffusion coefficient. Whereas a single Brownian particle trajectory is chaotic, averaging over many trajectories reveals a regular behavior.

The purpose of this article is to review recent efforts aiming at formulating a theory for anomalous diffusion processes-- i.e., those where  the mean square displacement does not follow Equation~(\ref{eq.sqr_disp}) which can be applied to many different situations from theoretical physics to biology.

In the next section, we introduce the pioneering works on diffusion, and  then  call attention to the existence of anomalous diffusion.  We discuss the main methods to treat anomalous diffusion and concentrate our efforts on the discussion of the generalized Langevin formalism.

%%%%%%%%%%%%%%%%%%%%%%%%%%%%%%%%%
\subsection{The tale of the three giants}

At the birth of the gravitational theory Isaac Newton mentioned  that he build up his theory based on the previous works of giants.  Also, one cannot talk about diffusion without underscoring the works of Albert Einstein,  Marian Smoluchowski, and Paul Langevin. At the dawn of the last century, the atomic theory was not widely accepted by the physical community. Einstein believed that the motion observed by Brown was due to the collisions of molecules such as proposed by Boltzmann in his famous equation. However, Boltzmann's equation was difficult to  solve and therefore he proposed a simpler analysis combining the kinetic theory of molecules with the Fick's law, from which he obtained the  diffusion equation 
 \begin{equation}
 \label{diff}
 \frac{\partial \rho(x,t)}{\partial t}=D\, \nabla^2\rho(x,t),
 \end{equation}
 where $\rho(x,t)$ is the density of particles at position $x$ and time $t$. The solution of the above equation yields $\rho(x,t)$ and
\begin{equation}
\lim _{t\rightarrow \infty }\langle  x^{2}(t)\rangle =\int x^2(t)\rho(x,t) dx= 2Dt,
\label{X2}
\end{equation}
where $\langle x(t)  \rangle =0$, assuming the symmetry $ \rho(-x,t) = \rho(x,t)$.
For simplicity, we have considered the motion one-dimensional and  set all  particles at the origin at time $t=0$. Generalization to two and three dimensions is straightforward. He then considered the molecules as single non-interacting spherical particles with radius $a$ and mass $m$, and subject to a friction $\gamma$ when moving in the liquid~\cite{Vainstein06,Einstein1905,Einstein56} .

Finally, he deduced the famous Einstein-Stokes relation \cite{Einstein56}  for the diffusion constant 
\begin{equation}
D=\frac{RT}{6\pi N_a a \eta}=\frac{RT}{mN_a\gamma}=\frac{RT\mu}{N_a},
\label{D}
\end{equation}
 where  $N_a$ is Avogadro's number, fundamental for atomic theory, but unknown at the time, $R$ is the gas constant, $\mu$ the mobility, $\eta$ the viscosity, and $T$ the temperature. 

  The scientific community became very excited and, in the years following Einstein's papers, some dedicated experiments   helped to verify  Equation~(\ref{X2}). Diffusion constants were measured and it became possible to estimate Avogadro's number from different experiments. Moreover, it was possible to estimate the radius of  molecules with few hundreds of atoms. Einstein was successful in demonstrating that atoms and molecules were not mere illusions as  critics used to suggest. Finally, the theory of Brownian motion  started to set  a firm ground. For instance, it became  possible to associate diffusion with conductivity in the case of a gas of charged particles. Supposing each has the same charge $e$ and is subjected to  a time dependent  electric field $\vec{E}=\vec{{E}}(\omega) \exp(-i\omega t)$, the   conductivity   $\tilde{\sigma}(\omega)$ can be defined by $\vec{J}(\omega)=\tilde{\sigma}(\omega)\vec{E}(\omega)$, where $\vec{J}(\omega)$ is the current. Now, it was possible to relate the diffusivity $\tilde{D}(w)$  with $\tilde{\sigma}(\omega)$  by the relation~\cite{Vainstein06,Dyre00,Oliveira05} 
 \begin{equation}
 \tilde{\sigma}(\omega)=\frac{ne^2}{k_BT}\widetilde{D}(\omega),
\label{Domega}
\end{equation}
where is  $n$ the carrier density.
From that it  was obvious that connections between a diversity of response functions could be obtained.
 
 Two  major achievements in the theory of stochastic motion were due to Smoluchowski. As expressed by Novak {\it et al.}~\cite{Nowak17}, ``One was the  Smoluchowski equation describing the motion of a diffusive particle in an external force field, known in the Western literature as the Fokker-Planck equation~\cite{Risken89, Salinas01}"
 \begin{equation}
 \label{FP}
 \begin{split}
 \frac{\partial P(v,t,v_0)}{\partial t}=- \frac{\partial}{\partial v} &\left[ A(v)P(v,t,v_0) \right] \\
 &+ \frac{1}{2} \frac{\partial^2}{\partial v^2} \left[ B(v)P(v,t,v_0) \right] ,
 \end{split}
 \end{equation}
 where
 \begin{equation}
 A(v)=\frac{1}{\delta t}\int_{-\infty}^{\infty}(v'-v)P(v',\delta t,v)\,dv'
 \end{equation}
 and
 \begin{equation}
 B(v)=\frac{1}{\delta t}\int_{-\infty}^{\infty}(v'-v)^2P(v',\delta t,v)\,dv'.
 \end{equation}
 Here   $P(v', \delta t,v)$  is the transition probability between two states with different velocities.
  The second one,  a fundamental cornerstone of molecular physical chemistry and  of cellular biochemistry \cite{Nowak17,Gadomski18}  is ``Smoluchowski's theory of diffusion limited coagulation of two colloidal particles."  Unfortunately, due to his premature death, the Nobel prize was not awarded to Smoluchowski. However, the scientific community pays tribute to him~\cite{Smoluchowski17}.
 
 The last of the three giants  was  Langevin, who
 considered Newton's second law of motion for a particle as \cite{Langevin08} 
\begin{equation}
m\frac{dv(t)}{dt}=-m\gamma v(t)+f(t),
\label{L}
\end{equation}
  dividing the environment's (thermal bath) influence  into two parts: a slow dissipative force, $-m\gamma v$,  with time scale $\tau= \gamma^{-1}$, and a fast random force $f(t)$, which changes in a time scale $\Delta t \ll  \tau $, subject to the conditions
 \begin{equation}
 \label{fmed}
 \langle  f(t)\rangle =0,
 \end{equation}
 \begin{equation}
 \label{fv}
  \langle  f(t)v(0)\rangle =0, 
 \end{equation} 
 and
 \begin{equation}
 \langle  f(t)f(t')\rangle= \Lambda \,\delta(t-t').
 \end{equation}
 If we solve Equation~(\ref{L}) and, using the equipartition theorem,  impose $\langle  v^2(t \rightarrow \infty) \rangle= \langle v^2\rangle_{eq}=k_BT/m$, where $k_B=R/N_a$ is the Boltzmann constant, we obtain $\Lambda =2m \gamma k_BT $ and write
\begin{equation}
\langle f(t)f(t')\rangle=2m \gamma k_BT\delta(t-t').
\label{FDT0}
\end{equation}
 This last equation establishes a relation between the fluctuation and the dissipation in the system reconnecting the useful, although artificial, separation of the two forces. This relation has been named the fluctuation-dissipation theorem (FDT) and is one the most important theorems of statistical physics. Equations (\ref{fmed}), (\ref{fv}), and (\ref{FDT0}) yield the velocity-velocity correlation function that reads~\cite{Reichl98} 
\begin{equation}
\label{Cv1}
   C_v(t)=  \langle v(t+t')v(t')\rangle=(k_BT/m)  \exp(-\gamma t),
\end{equation}
 the mean square displacement
\begin{equation}
\label{X22}
\langle x^2(t\gg \tau) \rangle= \int_0^tdt'\int_0^t dt''\langle v(t')v(t'')\rangle =2Dt,
\end{equation}
 and 
\begin{equation}
\label{Kubo}
 D=\int_0^{\infty}C_v(t)dt,
\end{equation}
known as the Kubo formula. This bring us back to  Einstein's results for diffusion, Equation~(\ref{D}).

  The simplification introduced by the Langevin formalism makes it easy to carry out analytical calculations and computer simulations. Consequently, the Langevin equation, and its generalization (Sec.  3),  has been applied successfully to the study of many different systems  such as chain dynamics~\cite{Toussaint04,Oliveira94,Oliveira96,Oliveira98a,Maroja01,Dias05}, liquids\cite{ Rahman62,Yulmetyev03}, ratchets \cite{Bao03a,Bao06}, and synchronization \cite{Longa96,Ciesla01}. However, its major importance was to relate fluctuation with dissipation.  

The Fokker Planck equation (\ref{FP}) with the Langevin choices  becomes
\begin{equation}
 \label{FP2}
 \frac{\partial P(v,t)}{\partial t}=\gamma\frac{\partial}{\partial v}\left[ v P(v,t)\right] +\gamma\frac{k_BT}{m} \frac{\partial^2}{\partial v^2}\left[ P(v,t)\right],
 \end{equation}
known as the Ornstein-Uhlenbeck equation. It obviously yields the same result as that of Einstein and Langevin.  This completes the tale of the three giants. Their work established the bases of non-equilibrium statistical mechanics opening a new field in research for the next decades. For example, it was demonstrated that  hydrodynamics could be obtained from the   Boltzmann equation in particular situations~\cite{Huang87},  so that the physics community could appreciate the work of yet another  giant. 

%%%%%%%%%%%%%%%%%%%%%%%%%%%%%%
\section{Breakdown of the normal diffusive regime}
\subsection{The different facets of the anomaly: subdiffusion and superdiffusion}

Inspection of different supposedly diffusive processes such as enhanced diffusion in the intracellular
medium \cite{Holek09}, cell
migration in monolayers~\cite{Palmieri15}, Levy flight search on a polymeric DNA \cite{Lomholt05}, or the Brownian motion in an inhomogeneous medium \cite{Durang15} reveals that the previous framework is not always fulfilled. Instead, in these and other similar examples the mean square displacement deviates from the linear temporal evolution. One of the most common anomalous behaviours is given by 
\begin{equation}
\label{X2anomalous}
 \lim_{t \rightarrow \infty} \left\langle r^2(t)\right\rangle  \sim t^\alpha,
\end{equation}
where $\alpha \neq 1$ is a real positive number \cite{Metzler00,Morgado02,Metzler04,Morgado04}.

\begin{figure*}[!ht]
\centering
\includegraphics[width=0.9\textwidth]{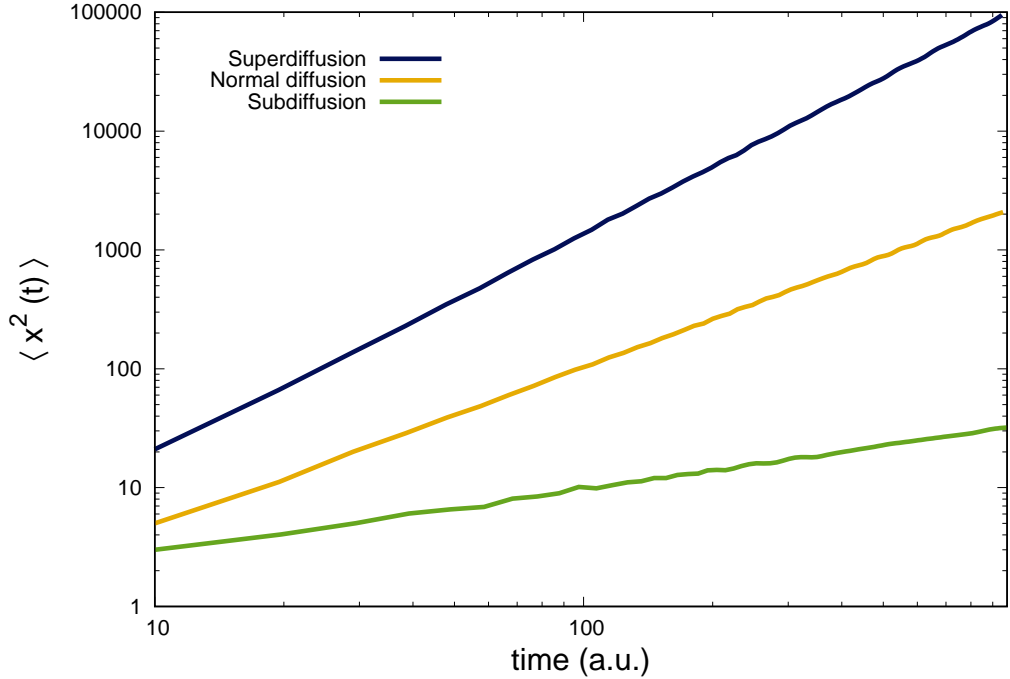}
\caption{(Color )  Time evolution of the mean square displacement  (in arbitrary units) for normal and anomalous diffusion. For the upper curve $\alpha=1.5$ (superdiffusion), for the middle curve $\alpha=1$ (normal diffusion)  and for the lower curve $\alpha=0.5$ (subdiffusion). From ref.~\cite{Lapas08}, with permission from Physical Review Letters. }
\label{fig:X2anomalous}
\end{figure*}

The origin of this discrepancy is the tacit assumption made in the derivation of (\ref{eq.sqr_disp}) that the Brownian particle moves in an infinite structureless medium acting as a heat bath. This assumption is generally incorrect when the Brownian motion takes place in a complex medium, as is the case of the previously mentioned examples.  We illustrate in  Fig. (\ref{fig:X2anomalous}) the mean square displacement $\langle x^ 2(t) \rangle$ as a function of $t$ for three distinct  Brownian motions in one dimension. From the upper curve downwards,  we have $\alpha=1.5$  (superdiffusion),  $\alpha=1$ (normal diffusion), and $\alpha=0.5$ (subdiffusion).

The recent interest in the study of complex systems has led to an increased focus on anomalous diffusion~\cite{Metzler00, Morgado02, Metzler04,Sancho04, Yulmetyev03,Holek09,Durang15}, which has been described   in biological systems~\cite{Holek09,Palmieri15,Lomholt05},   protostellar birth~\cite{Vaytet18},  complex fluids~\cite{Mason95,Grmela97,Bakk02,Sehnem2014,Sehnem15,cabreira18}, electronic transportation~\cite{deBrito95,Monte00,Monte02,Kumakura05,Borges06,Novak05}, porous media and infiltration \cite{Filipovitch16,Reis16,Reis18},  drug delivery~\cite{GomesFilho16,Ignacio17,Gun17,Soares17},  fractal structures and networks~\cite{Mandelbrot82,Stauffer95,Cristea14,Balankin17,Balankin18},  and in  water's anomalous behavior~\cite{Barbosa11,Bertolazzo15,Silva15},  phase transitions in synchronizing oscillators \cite{Ciesla01,Longa96,Bier16,Pinto16,Pinto17}, to name a few.

There are many formalisms that describe anomalous diffusion, ranging from thermodynamics~\cite{Perez-Madrid04a,Rubi88,kusmierz18}, and fractional derivatives~\cite{Metzler00,Metzler04} to generalized Langevin equations  (GLE)~\cite{Hanggi90,Morgado07,Lapas07}. The goal of this short review is to call attention to relevant research within the GLE framework and to some fundamental theorems in statistical mechanics.

%%%%%%%%%%%%%%%%%%%%%%%%%%%%%%
\section{The generalized Langevin equation approach}
%%%%%%%%%%%%
\subsection{Non-Markovian processes}
 The Langevin formalism within its classical description has some  restrictions: 
\begin{enumerate}
\item It has a relaxation time $\tau=\gamma^{-1}$ , and an unspecified $\Delta t \ll \tau $, while a diffusive process in a real system can present several time scales;
\item  For short times $t < \Delta t$, predictions are unrealistic; for example, the derivative of $C_v(t)$ is zero \cite{Lee83} at $t=0$, while in  Langevin's formalism $C_v(t)=\exp(-\gamma  | t|  )$,  exhibiting a discontinuity.
\item If no external field is applied it cannot  predict anomalous diffusion.
\end{enumerate}

Up to now we have used time  and ensemble averages without distinction, i.e. we implicitly used the Boltzmann Ergodic hypothesis (EH).   The ensemble average for a variable $B(t)$ is defined by
\begin{equation}
\label{Bens}
 \langle B(t)  \rangle= \int \exp{(- E/(k_BT))}\Omega(E,B)B(t)dE, 
\end{equation}where $\Omega(E,B)$ is the number of states for a given energy $E$.  For the time average we have
\begin{equation}
\overline{B(t)}=\frac{1}{\tau_0}\int_{-\tau_0/2}^{\tau_0/2}B(t+t')dt'.
\label{time}
\end{equation} 
 For times $\tau_0  \gg \tau$, the Ergodic Hypothesis (EH)  reads
\begin{equation}
\label{EH}
\overline{B(t)} = \langle B(t)  \rangle, 
\end{equation}which means that the system should be able to reach every accessible state in configuration space given enough time.  This is expected to be true for equilibrated macroscopic systems and also for systems that suffer small perturbations close to equilibrium.  

 In this section, we generalize  Langevin`s equation and study some of its consequence for diffusion. The first correction to the FDT was done by Nyquist~\cite{Nyquist28} who formulated a quantum version of the FDT.
 Later on Mori~\cite{Mori65,Mori65} and Kubo~\cite{Kubo74} used a projection operator method to  obtain the equation of motion,  the Generalized Langevin equation (GLE),  
\begin{equation}
\label{GLE}
\frac{d A(t)}{d t}=-\int _{0}^{t}\Pi (t-t')A(t')\,d t'+F(t),
\end{equation}
for a dynamical operator  $A(t)$, where $\Pi (t)$ is a non-Markovian  memory  and $F(t)$ is a random variable subject to 
\begin{enumerate}
\item  $\overline{ F(t) } =0$,
\item  $\overline{ F(t)A(0)}  =0 $, and, 
\item  the Kubo fluctuation-dissipation theorem (FDT)~\cite{Kubo91,Kubo66} 
\begin{equation}
\label{fdt}
C_F(t-t')=\overline{ F(t)F(t')}  = \langle A^2 \rangle_{eq} \, \Pi(t-t').
\end{equation}
\end{enumerate}

  In this way, it is clear that time translational invariance holds in the Kubo formalism.  An alternative to  the projection operators, the  recurrence relation method, was derived by Lee~\cite{Lee83,Lee82,Lee83a,Lee84}.

 It is easy to show that  Equation (\ref{GLE}) can give rise both to normal and to  anomalous diffusion~\cite{Vainstein06}.  Let us consider two limiting case examples:  first, when $\Pi(t)=2\gamma \delta(t)$, we recover the normal Langevin equation  (\ref{L})  with normal diffusion ($\alpha=1$); second,  for a constant memory $\Pi(t)=K$, Equation~(\ref{GLE}) becomes 
\begin{equation}
\frac{d^2x}{dt^2}=-Kx +F(t),
\label{harm_osc}
\end{equation}
where
\begin{equation}
\label{x}
x(t)=\int_0^t A(t')\,d t'.
\end{equation}
Equation~(\ref{harm_osc}) is the equation of motion for a harmonic oscillator with zero diffusion constant ($D=0$) and with exponent $\alpha=0$. Consequently,  we may have different classes of diffusion for distinct memories.

Now, we can study the asymptotic behavior of Equation (\ref{x}) or of  its second moment, Equation~(\ref{X2anomalous}), to characterize the type of diffusion presented by the system for any memory $\Pi(t)$.   Much information about the system's relaxation properties can be obtained by studying the correlation function
\begin{equation}
\label{CA}
C_A(t,t')=\overline{ A(t)A(t')}= C_A(t-t').
\end{equation}
The existence of stationary states warrants time-translation invariance so that the the two-time correlation function becomes  a function only of the difference between two times, such as in the Kubo FDT above.   We shall return to this point later. 
 The main equation we are interested in  is Equation  (\ref{eq.sqr_disp}), where for anomalous diffusion, $D$ should be replaced by $D(t)$ now defined  by  
$D(t)= \int_0^{t}C_A(t')dt'$
then
\begin{equation}
\label{Kubo2}
 \lim_{t \rightarrow \infty}D(t)=\lim_{t \rightarrow \infty} \int_0^{t}C_A(t')dt'=\lim_{z \rightarrow 0} \widetilde{C}_A(z).
\end{equation}
From here onwards, the tilde over the function stands for the Laplace transform. For the last equality, we use the final-value theorem for Laplace transforms~\cite{Gluskin03}, which  states that if $f(t)$ is bounded on $(0,\infty )$ and   $\lim_{t \rightarrow \infty} f ( t ) $ has a finite limit, then $ \lim _{t\rightarrow \infty }f(t) = \lim_{z\rightarrow 0}{z \widetilde{f}(z)}$.  Also note   that the Laplace transform of the integral of a function is the Laplace transform of the function divided by $z$, then $\widetilde{D}(z)=\widetilde{C}_A(z)/z$. 
Using $z \propto 1/t$,  we obtain the asymptotic behavior of $D(t)$. In order to do that,   we multiply Eq. ( \ref{GLE}) by $A(0)$, take the average and use the  conditions $(1)$ and $(2)$ above for the noise  to obtain the self-consistent equation
\begin{equation}
\label{self_consistent}
\frac{d R(t)}{d t}=-\int_0^t \Pi(t-t')R(t')\,d t',
\end{equation}
where for  simplicity we have defined 
\begin{equation}
\label{cor}
R(t)=\frac{C_A(t)}{C_A(0)}.
\end{equation}
Note that from Equation (\ref{GLE}), we need to average from a large number of stochastic trajectories, while to obtain Eq. ~(\ref{cor}), it is only necessary to solve a single equation, i.e. Eq. (\ref{self_consistent}). Further insight can be gained by analyzing the Laplace transformed version of Equation~(\ref{self_consistent}) 
\begin{equation}
\label{laplace_R}
\widetilde{R}(z)=\frac{1}{z+\widetilde{\Pi }(z)}.
\end{equation}
From here, it is clear that the knowledge of $\widetilde{\Pi}(z)$ in the limit $z\to 0$ completely defines the asymptotic dynamics. For instance, if  $\lim_{z \rightarrow 0} \widetilde{\Pi}(z) \propto z^\mu$ then Equation~(\ref{Kubo2}) becomes
\begin{equation}
\label{diff_beta}
\lim_{t \rightarrow \infty}D(t) \propto t^\beta 
\end{equation}
and consequently  \cite{Morgado02}
\begin{equation}
\label{alphaeq}
\alpha=\beta+1,
\end{equation}
where
\begin{equation}
\label{expbeta}
\beta= 
\begin{cases}
\mu , &\text{ if~~ } -1<\mu <1\\
  1,  &\text{ if~~ } \mu  \geq 1,\\
\end{cases}
\end{equation}
We see that we have a cutoff  limit for the exponent $\beta$  and, therefore, also for $\alpha$.

 Due to the existence of correlations in the GLE that arise through hydrodynamical interactions~\cite{Reichl98}, it has been proposed~\cite{Morgado02,Vainstein05,Morgado04} to establish a connection between the random force $F(t)$  and the  noise density of states $\rho (\omega )$ of the surrounding media, modeled as a thermal bath  of harmonic oscillators~\cite{Morgado02, Hanggi90} of the form   \begin{equation}
 \label{Noise}
 F(t)= \int C (\omega )\cos [\omega t+\phi(\omega)]d \omega,
 \end{equation}
 where  $0 < \phi < \pi$ are random phases.
  Now, using the Kubo FDT, eq. (\ref{fdt}), and time averaging over the cosines, we obtain the memory as~\cite{Morgado02,Costa03}
\begin{equation}
\label{memory}
\Pi (t)=\int \rho (\omega )\cos (\omega t)d \omega,
\end{equation}
 which is an even function independent of the noise distribution.  Here,  $\rho(\omega)=C^2 (\omega )\langle A^2 \rangle_{eq}/2$.

 The consequences of considering a colored noise given  by a generalization of the Debye spectrum
\begin{equation}
\label{noise_dos}
\rho (\omega )= 
\begin{cases}
\frac{2\gamma }{ \pi } \left( \frac{ \omega }{\omega_s} \right )^\nu, & \text{ if } \omega<\omega_s\\
0, &\text{ otherwise},
\end{cases}
\end{equation}
with $\omega_s$ as a Debye cutoff frequency were analyzed in detail in~\cite{Vainstein06a}. The reason for the choice of this functional form for the noise density of states is that it was previously shown in~\cite{Morgado02,Costa03} that if  $\widetilde{\Pi}(z) \propto z^\mu$ as $z\rightarrow 0$,  then the same restriction as in Equation (\ref{expbeta}), with $\nu=\mu$, applies and the diffusion exponent  \cite{Morgado02}  is given by  Equation (\ref{alphaeq}). 

Later, this problem was revisited by Ferreira \textit{et al.}~\cite{Ferreira12} in which a generalized version of  Equation (\ref{X2anomalous}) was considered, namely  
\begin{equation}
\label{X2anomalous2}
 \lim_{t \rightarrow \infty} \left\langle r^2(t)\right\rangle  
  \sim  t^{\alpha}(ln(t))^{ \pm n}.
\end{equation}
  
 Most authors~\cite{Metzler00,Morgado02,Metzler04,Morgado04} have reported the cases  of  anomalous diffusion where, $n=0$ and $\alpha \neq 1 $. 
However, some authors such as for example, Srokowksi~\cite{Srokowski00,Srokowski13} reports situations were for $t \rightarrow \infty $, the dispersion behaves as
\begin{equation}
\langle x^2(t) \rangle \propto t/\ln(t), 
\end{equation}i.e.,  a weak subdiffusive behavior for which we can say that $\alpha = 1^-$. In this way  Ferreira \textit{et al.} \cite{Ferreira12} generalizes the concept of $\alpha$, to associate with Eq. (\ref{X2anomalous2}), the $ \alpha^{\pm}$ exponents, which  
 arise analogously to the critical exponents of a phase transition~\cite{kadanoff67,kadanoff00,Kenna06}. 
For example,  in magnetic systems with temperatures $T$ close to the transition temperature $T_c$,  the specific heat at zero field, $H=0$, exhibits the power law behavior $C_{H=0} \propto|T-T_c|^{-\alpha}$, where  $\alpha$ is the critical exponent.  However, for the two-dimensional Ising model \cite{kadanoff67} the critical exponent can be considered  $\alpha = 0^{+}$, since  the specific heat behaves logarithmically, $ C_{H=0} \propto  \ln{|T-T_c|}$, instead.    Logarithmic corrections~\cite{Kenna06} to scaling have also been applied to the diluted Ising model in two dimensions in~\cite{Kenna08}.   This generalized nomenclature is pertinent  because there are many possible combinations of both logarithmic and power law behaviors. 
 This result highlights the existence of different types of diffusion.

  In this way, for the density of states  (\ref{noise_dos}),  the generalized $\alpha$ becomes 
\begin{equation}
\label{diff_exponent}
\alpha=
\begin{cases}
2, &\text{ if~~ } \nu>1\\
2^-, &\text{ if~~ } \nu=1\\
1+\nu, &\text{ if~~ } -1<\nu<1.
\end{cases}
\end{equation}
In Equation (\ref{noise_dos}), we choose the constant such that for normal diffusion $\widetilde{\Gamma}(z=0)=\gamma$.  The  exponent for Ballistic diffusion (BD), $\alpha=2$, is the maximum for diffusion in the absence of an external field. The slow ballistic motion $\alpha=2^{-}$ has properties that differ markedly from the ballistic case, see sections (3.3 and 3.4).

\subsection{Non-exponential relaxation}

Besides the importance of the asymptotic behavior, the study of the correlation $R(t)$ for finite times is also obviously significant,  and there exists a  vast literature describing non-exponential behavior of correlation functions in systems ranging from plasmas to hydrated proteins~\cite{Rubi04,Santamaria-Holek04,Vainstein03a,Santos00,Benmouna01,Peyrard01,Colaiori01,Ferreira91,Bouchaud91}, since the pioneering works of Rudolph Kohlrausch~\cite{Kohlrausch54} who described charge relaxation in Leyden jars using stretched exponentials, $ R(t) \approx  \exp{[-(t/\tau )^{\beta} ]} $ with $0 < \beta < 1$, and his son Friedrich Kohlrausch \cite{Kohlrausch63}, who observed two universal behaviors: the stretched exponential and the power law.  Since many features are shared among such systems and those that present anomalous diffusion~\cite{Vainstein05,Lapas15}, it is natural that similar methods of analysis can be applied to both.
 For example,   from Equation (\ref{self_consistent}),  $\frac{dR(t)}{dt}$  must be zero at $t=0$, which is at odds with the result $R(t)=\exp(-\gamma \vert t\vert )$  of  the memoryless Langevin equation. Nevertheless, we know that the exponential can be a reasonable approximation in some cases: Vainstein  {\it et al.}~\cite{Vainstein06a} have presented  a large diversity of correlation functions that can be obtained from Equation~(\ref{self_consistent}) once  $\Pi(t)$ is known. Since,
from Equation  (\ref{memory}),  $\Pi(t)$  is an even function then we can write 
\begin{equation}
\label{memory2}
\Pi (t)=\sum_{n=0}^{\infty}b_n t^{2n}.
\end{equation}
From Equation (\ref{self_consistent}), they proved that $R(t)$ must also be an even function, therefore
\begin{equation}
\label{R2}
R(t)=\sum_{n=0}^{\infty}a_n t^{2n},
\end{equation}
with $a_0 = R(0) = 1$.  We insert Equations  (\ref{memory2}) and (\ref{R2})  into Equation~(\ref{self_consistent}) to obtain the recurrence relation~\cite{Vainstein06a}
\begin{equation}
a_n=-\frac{2\gamma \omega_s}{\pi(2n)!}\sum_{l=0}^{n-1}\frac{(-1)^l[2(n-1-l)]!\, \omega_s^{2l}}{(2l+1+\nu)} a_{n-1-l}, 
\end{equation}
which  shows the richness and complexity of behavior that can arise from a non-Markovian model. The above defined convergent power series represents  a large  class of functions, including the Mittag-Leffler function~\cite{Mittag-Leffler05} which behaves as a stretched exponential for short times and as an inverse power law in  the long  time scale.  Note that even for the simplest case of normal diffusion, $\nu=0$, $R(t)$ is not an exponential since at the origin its derivative is zero; however, for broad-band noise  $\omega_s \gg  \gamma $, i.e.,  in the limit of white noise it approaches the exponential  $R(t)= \exp(-\gamma t)$, for times larger than $\tau_s= \omega_s^{-1}$.

\begin{figure*}[!ht]
\centering
\includegraphics[width=0.5\textwidth,angle=270]{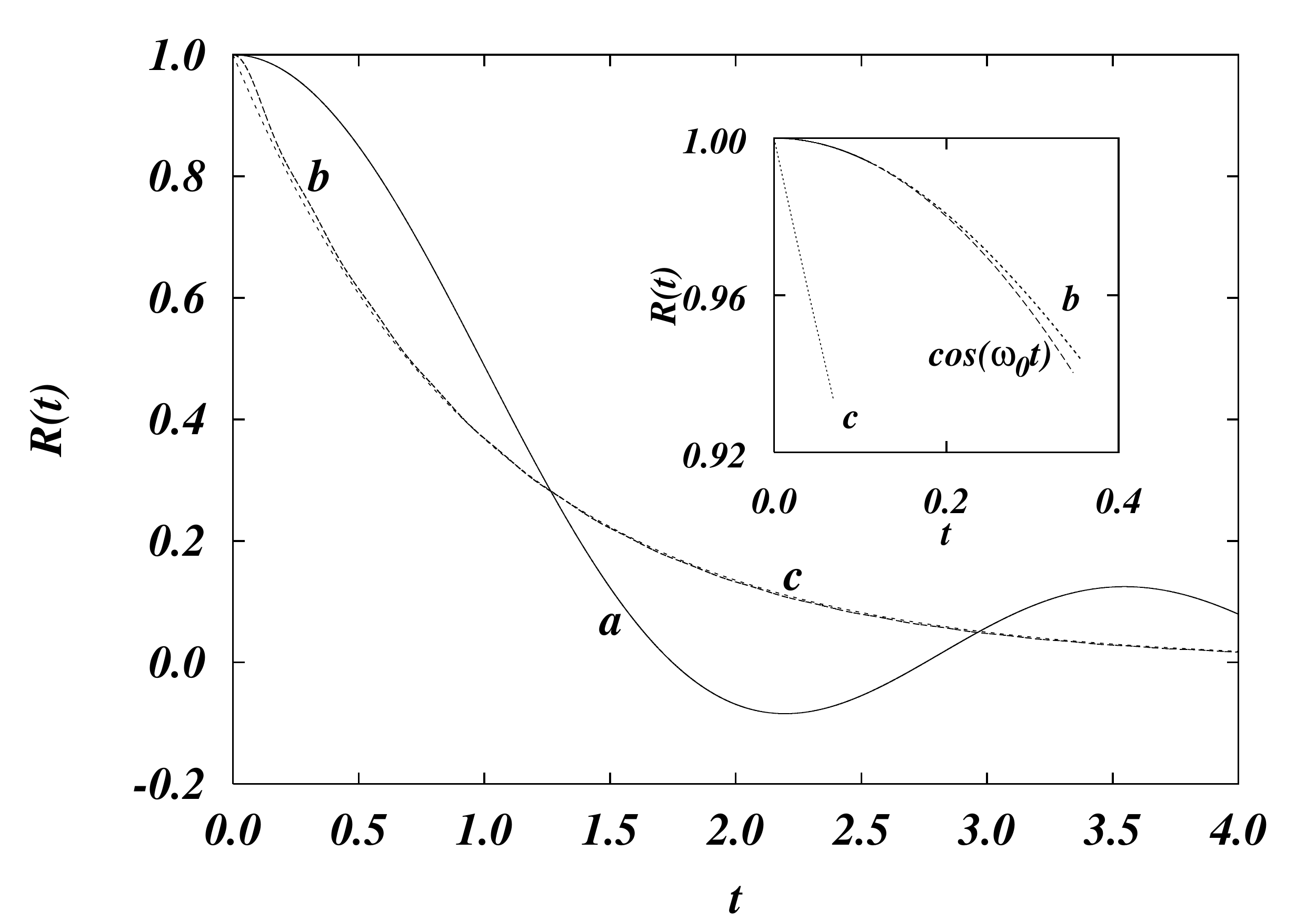}
\caption{Plot of the correlation function $R(t)$ for normal diffusion ($\nu=0$). Here, $\gamma=1$ and $\omega_s=2 $ and $20$, for curves $a$ and $b$, respectively. Curve $c$ is the plot of $\exp(-\gamma t)$. Note that curves $b$ and $c$ are indistinguishable, except near the origin (see inset). Inset: Plot of curves $b$, $c$ and $\cos(\omega_0 t)$, near the origin. From ref.~\cite{Vainstein06},  with permission from Europhysics Letters. } 
\label{fig:correlation_anomalous}
\end{figure*}

 In Fig.~(\ref{fig:correlation_anomalous}), from~\cite{Vainstein06a}, we display the rich behavior of the  correlation function $R(t)$  for the case of normal diffusion.  Here, $\gamma=1$ and $\omega_s=2 $ and $20$, for curves $a$ and $b$, respectively. Curve $c$ is the plot of $\exp(-\gamma t)$. In the inset, we highlight that although curve $b$ and the exponential approach one another for long times, for short times they differ appreciably.  Also plotted is $\cos(\omega_0 t)$, with $\omega_0 = \sqrt{\Pi(0)}$.

%%%%%%%%%%%%%%%%%%%%%%%%%%%%%%
\subsection{Basic theorems of statistical mechanics}
\subsubsection{The decay towards an equilibrium state}
One of the most important aspects of  dynamics is to observe the asymptotic behavior of a system or how it approaches   equilibrium (or not). Observe that a direct solution for Equation~(\ref{GLE}) is
\begin{equation}
A(t)=A(0)R(t)+\int_0^tR\left(t-t'\right)F(t)dt\label{eq:sol_gle}.
\end{equation}
Given the initial states $A(0)$, it is possible to average over many trajectories to obtain the temporal evolution of the moments $\overline{ A^{n}(t)}$, with $n=1,2,\ldots$.   The first moment arises directly from an  average of Equation~(\ref{eq:sol_gle}),
\begin{equation}
\overline{ A(t)}=\overline{ A(0)} R(t)\text{.}\label{p_ave}
\end{equation}
 Taking the square of Equation~(\ref{eq:sol_gle}) and averaging, we obtain
\begin{equation}
\overline{  A^2}=\overline{ A^2}+R^2(t)\left[\overline{ A^2(0)}-\langle A^{2}\rangle_{eq}\right]\text{.}\label{p2_ave}
\end{equation}
The skewness is defined as a measure of the degree of asymmetry of the distribution of $A(t)$, and is given by \cite{Lapas07}
\begin{equation}
\label{ske}
\zeta(t)=\left[\frac{\sigma_A(0)}{\sigma_A(t)} \right]\zeta(0)R^3(t),
\end{equation}
where $\sigma_A(t)=\overline{ A^2(t)}-\overline{A(t)}^2$.  
We also obtain  the  non-Gaussian factor \cite{Lapas07}
\begin{equation}
\label{NG}
\eta(t)=\left[\frac{\overline{ A^2(0)}}{\overline{ A^2(t)}} \right]\eta(0)R^4(t).
\end{equation}
Consequently, we see that $R(t)$ completely determines these averages.  One can note that if the system is originally at equilibrium   $\overline{A(0)}=0$ and $\overline{ A^2(0)} =\left\langle  A^2\right\rangle_{eq}$, then the system remains in equilibrium. If the system is not in equilibrium, then
\begin{equation}
\lim_{t \rightarrow \infty }R(t)=0
\label{MC}
\end{equation}
drives the system  towards equilibrium.  The condition  stated in Equation (\ref{MC}) is called the mixing condition (MC) and is a fundamental concept in statistical mechanics,  which asserts that after a long time the system reaches  equilibrium and forgets all initial conditions.

  Note that parity is conserved as well: if the initial skewness is null, $\zeta(0)=0$ in Eq. (\ref{ske}), it will remain null during the whole evolution;  the same holds for the nongaussian factor.

\subsubsection{The Kinchin theorem and ergodicity}
 The Khinchin theorem (KT)~\cite{Lapas08,Khinchin49} states that if the the MC holds, then ergodicity holds. As shown below, anomalous diffusion  is a good field for testing ergodicity breaking~\cite{Lapas08, Weron10,Lapas07}.
 For a situation where the MC is violated, we have
\begin{equation}
\kappa= \lim_{t \rightarrow \infty }R(t)=\lim_{z \rightarrow 0}z\widetilde{R}(z)=\lim_{z \rightarrow 0}\left[1+\frac{\widetilde{\Pi}(z)}{z} \right]^{-1}\neq 0,
\label{MCV}
\end{equation}
 where $\kappa$ is the nonergodic factor \cite{Costa03,Lapas08,Bao06,Bao05a}.  For example  for $\widetilde{\Pi}(z\rightarrow 0)\propto bz$ ,  $ \alpha=2$, and 
\begin{equation}
\kappa = \frac{1}{1+b}.
\label{MCV2}
\end{equation}
I.e., the MC is violated in the ballistic motion, and consequently ergodicity is violated, but not the KT.  Note that for $\alpha=2^{-}$,   where $R(t \rightarrow \infty) \rightarrow 1/\ln(t) \rightarrow 0$, the MC is not violated. In this way the MC is satisfied for all diffusive processes in the range $0<\alpha<2^{-}$ \cite{Ferreira12}.

It is interesting to observe that Equation~(\ref{MCV}) for long time behavior is equivalent to the condition
\begin{equation}
\Lambda=\int U(\vec{r})\,d\vec{r}\to \infty,
\end{equation}
for systems with long range interactions \cite{Silvestre,Campa09}, where  $U(\vec{r})$ is the potential between the particles and the integration is performed over all space.

\subsubsection{Gaussianization}

\begin{figure*}[!ht]
\centering
\includegraphics[width=0.3\textwidth]{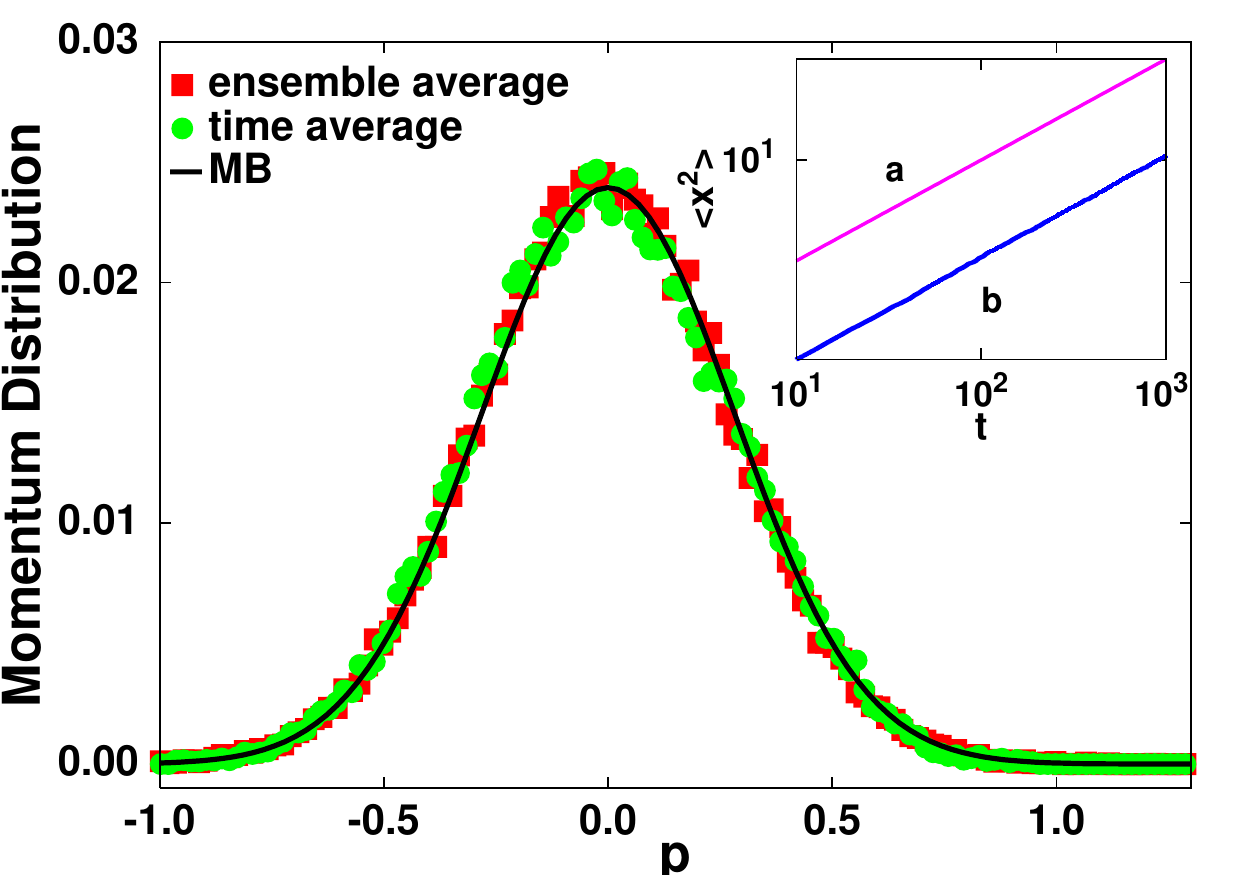}
\includegraphics[width=0.3\textwidth]{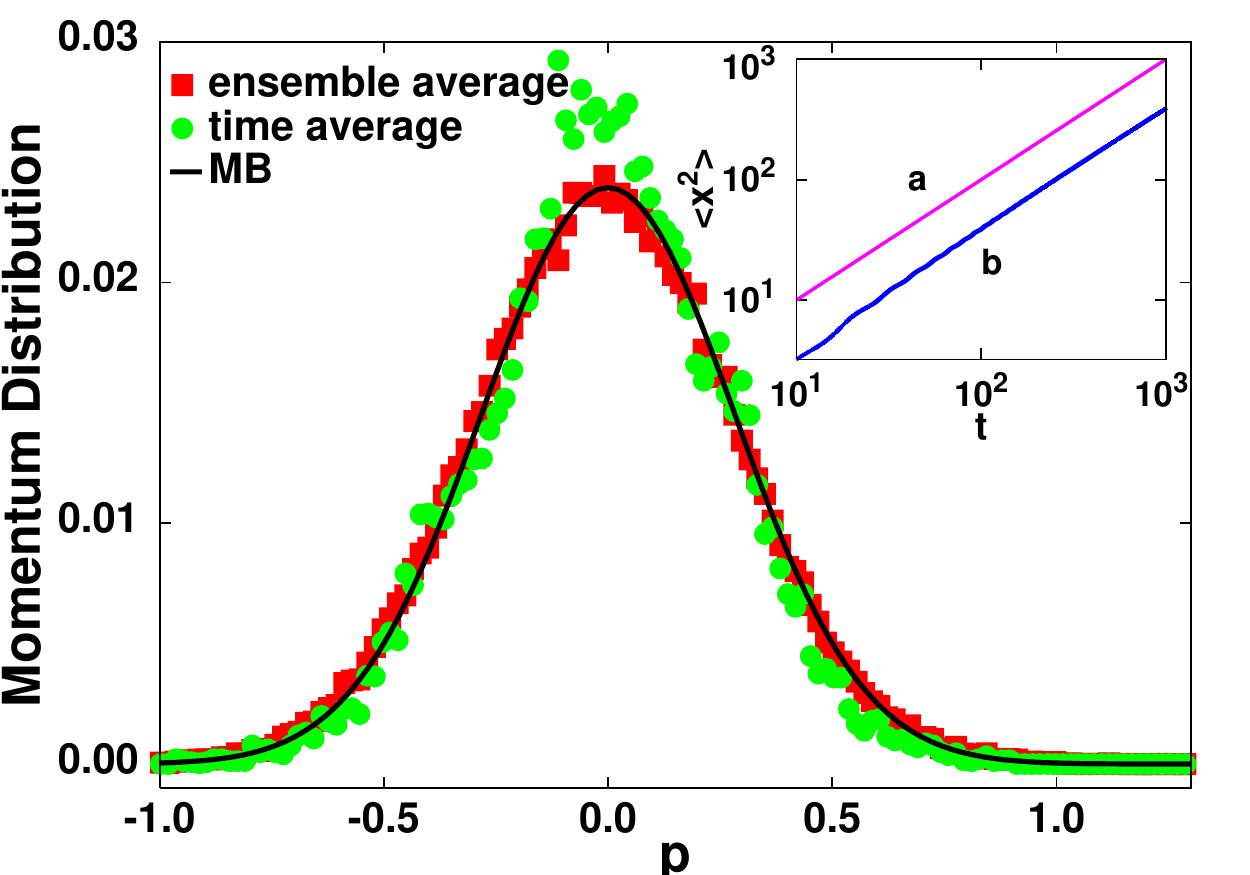}
\includegraphics[width=0.3\textwidth]{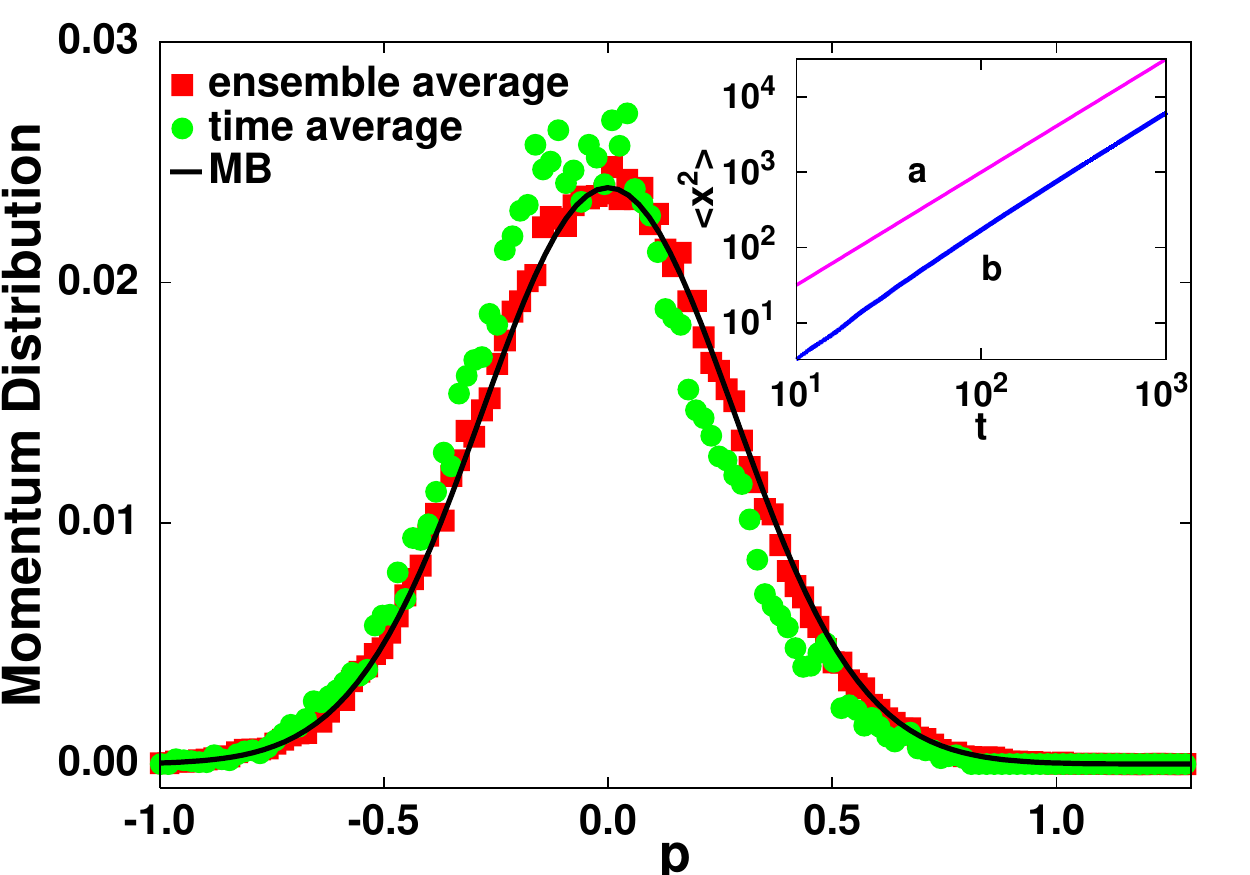}
\caption{(Color online) Numerical results for the probability distribution function for subdiffusion (left, $\alpha=0.5$), normal diffusion (middle, $\alpha=1$) and superdiffusion (right, $\alpha=1.5$). The time averages (circles) are obtained by following one particle trajectory and calculating the histogram for times from $t=100$ to $t=5000$. For the ensemble averages (squares), we calculate the histogram using $5\cdot10^{4}$ particles, at time $t=1000$. The continuous line is the Maxwell-Boltzmann distribution. Insets: Curves $a$ correspond to the functions $t^\alpha$ and curves $b$ to the simulated mean square displacements. From Lapas {\it et al.} \cite{Lapas08},  with permission from Physical Review Letters.}
\label{fig:hist}
\end{figure*}

As an illustration of the analytical results, we numerically integrated the GLE, Equation~(\ref{GLE}), to approximate the particle velocity distribution function, using  Equations (\ref{memory}) and (\ref{noise_dos})  to generate the memory for $\nu=-0.5$, $0$, and $0.5$, which by Equation~(\ref{diff_exponent}) give $\alpha = 0.5$, $1$, and $1.5$, respectively.  The results are exhibited in Fig.~(\ref{fig:hist}), from ~\cite{Lapas08},  where we show the 
 probability distribution functions as function of the momentum $p$. From left to right, we have  subdiffusion ($\nu=-0.5$), normal diffusion ($\nu=0$), and superdiffusion ($\nu=0.5$) for the values $a=\frac{2\gamma}{\pi}=0.25$. A value $\omega_s=0.5$ was used for normal and superdiffusion. In the case of subdiffusion,  a broader noise $\omega_s=2$ is needed for it to arrive at  the stationary state. It is expected  that $R(t\rightarrow\infty)=0$  in all cases, and that the EH will be valid even for the subdiffusion (superdiffusion). It should be noted that  despite large fluctuations in the time average, there is a good agreement between the ensemble and time distributions,   in agreement with Eqs.  (\ref{p_ave}) to (\ref{MC}), indicating the validity of the EH. In all cases, the distributions converge to the expected Maxwell-Boltzmann distribution, in accordance with analytical results~\cite{Lapas07}.

\begin{figure*}[!ht]
\centering
\includegraphics[width=0.8\textwidth]{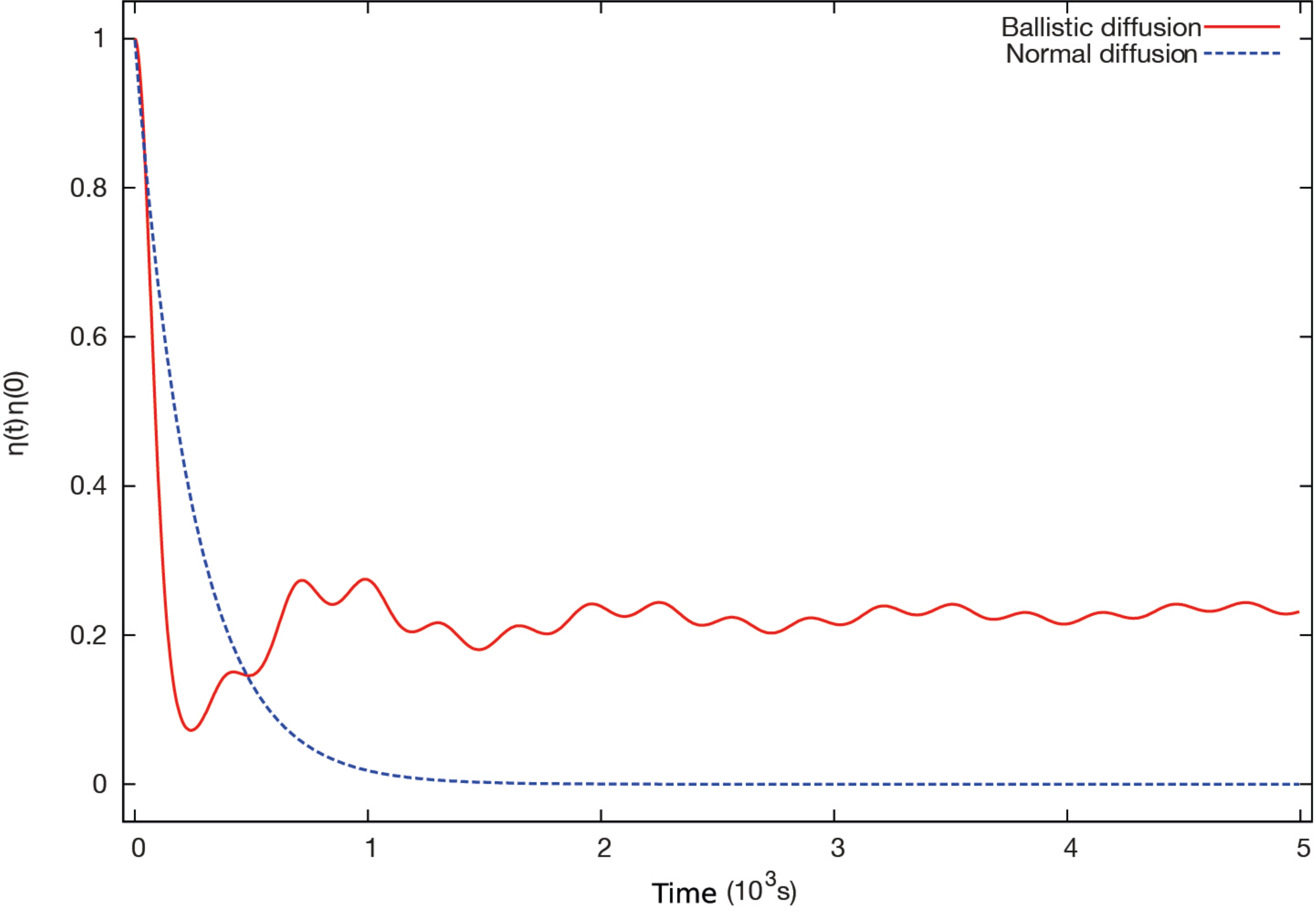}
\caption{(Color online)  Time evolution of the normalized non-Gaussian factor $\eta(t)/\eta(0)$ for normal and ballistic diffusion. In both cases, we consider the initial PDF as a Laplace distribution with unit mean and unit variance.  The functions $R(t)$ are as in Fig. (\ref{fig:ekt}).  From \cite{Lapas07}, with permission from Europhysics Letters. }
\label{fig:ngaussian}
\end{figure*}

 In Fig. (\ref{fig:ngaussian} we show  the evolution of the nongaussian factor.  We see that for the Ballistic diffusion (BD), 
 it does not reach a null value, but it evolves towards it.   Note that even for a situation where $\kappa \neq 0$, the nongaussian factor will be very small and the probability of it being  non-zero in  simulations after a long time is very small.  For example for $\kappa=0.1$,  there will a factor of $10^{-4}$ in relation (\ref{NG}).

\begin{figure*}[!ht]
\centering
\includegraphics[width=0.8\textwidth]{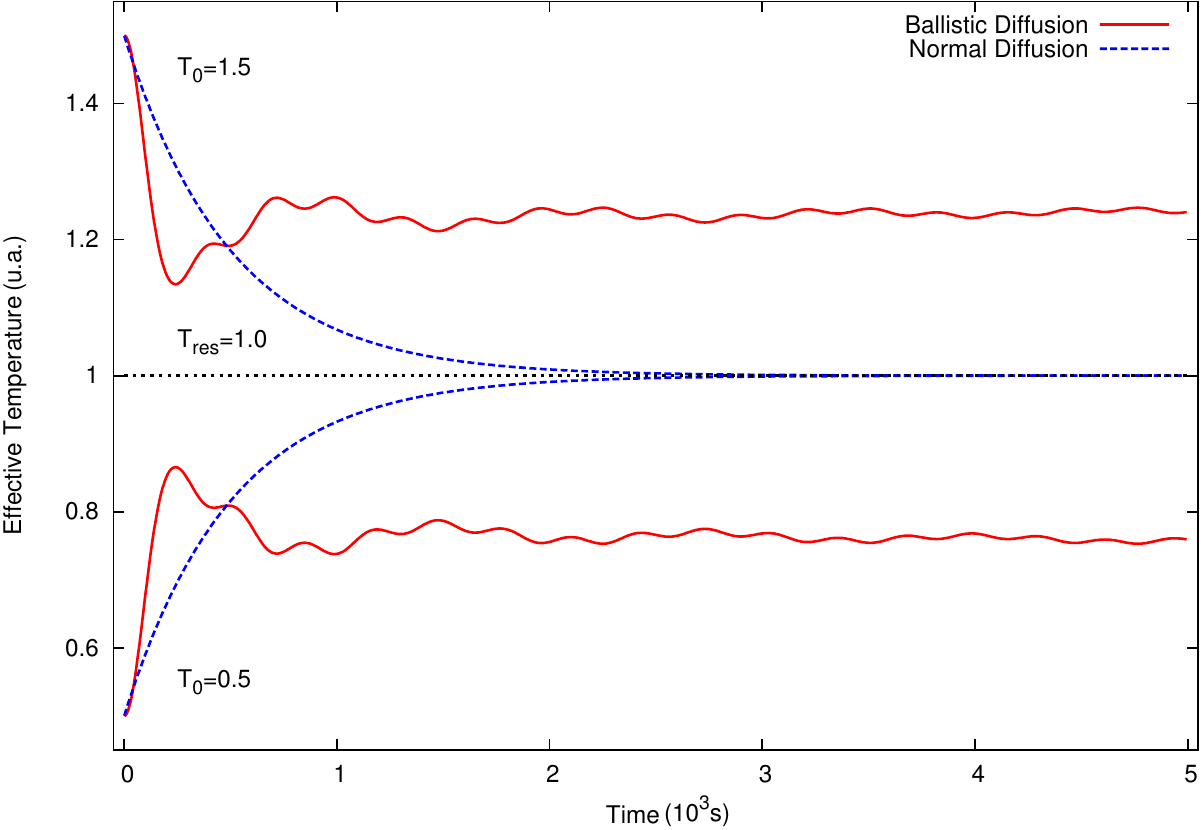}
\caption{(Color )  Time evolution of the effective kinetic temperature (in arbitrary units) for normal and ballistic diffusion. For BD we use the noise distribution $\rho(\omega) = 2\gamma /\pi$ for $1 < \omega < 4$  and $\rho(\omega) = 0$, otherwise, and we found $R(t)$ numerically (solid curve).  Since BD converges very slowly, we use $\gamma = 1$ for BD and $\gamma = 10^{-3}$ for normal diffusion with $R(t)= \exp(-\gamma t)$, so we can compare both. We choose $T_0 = 1.5$ and $T_0 = 0.5$ for the initial temperatures. In both cases we used $T=1$ for the reservoir temperature. From \cite{Lapas07},  with permission from Europhysics Letters.  }
\label{fig:ekt}
\end{figure*}

In Fig. (\ref{fig:ekt}), from~\cite{Lapas07}, we show the evolution of the kinetic temperature of the system by taking $A^2(t)$, with  $A=P$,  the momentum of the particles. In this case we should have $ \langle A^2 \rangle_{eq} -\langle A \rangle^2_{eq} \propto T$~\cite{Lapas07} and the temperature evolution can be obtained from Equation~(\ref{p2_ave}). We consider both normal and ballistic diffusion (BD) in a reservoir characterized by $T=1.0$ with initial high and low temperatures $T_0 = 1.5$ and $T_0 = 0.5$, respectively.	For normal diffusion (dashed curve), we have $R(t) = \exp{(-\gamma t)}$, with $\gamma = 10^{-3}$. For BD, $R(t)$  is calculated numerically in \cite{Lapas07}  with $\gamma = 1$. Since BD's relaxation is slow, we take  a “friction” $\gamma$ a thousand times larger than that of normal diffusion  for comparison. As expected, in the case of normal diffusion the system's temperature always relaxes to that of the reservoir, while for BD the temperature approaches that of the reservoir without reaching it~\cite{Lapas07,Lapas15}.  Figure~(\ref{fig:ngaussian}) displays the normalized non-Gaussian factor, Equation (\ref{NG}), as
a function of time \cite{Lapas07}  for the cases in the previous figure, with the same convention for the labeling.   For normal diffusion, the system's probability distribution evolves towards a Gaussian, which is not the case for BD. In the latter case, $R(t)$ oscillates around the value predicted by Equation~(\ref{MCV2}), even for long times,  In both cases, the initial probability distribution function  was the Laplace distribution, with $\langle A^2(0)\rangle=1$ and $\langle A(0)\rangle=0$.

 \subsubsection{The fluctuation-dissipation theorem}
 
Figures  (\ref{fig:hist})--(\ref{fig:ngaussian}) are just illustrations of the results that can be obtained analytically from Equations (\ref{MC}) to (\ref{EH}).  Lapas {\it et al.}~\cite{Lapas08,Lapas07} have shown that the  KT is valid for all forms of diffusion, and that the ballistic diffusion violates the EH, but not the KT.

 Although it is expected that a system in contact with a heat reservoir  will be driven to equilibrium by fluctuations, Figs.(\ref{fig:ekt}) and (\ref{fig:ngaussian}) show that it is not always the case and in some far from equilibrium situation it may not happen. The concept of ``far from equilibrium'' is itself sometimes misleading, since it depends not only on the initial conditions, but also on the possible trajectories the system may follow~\cite{Costa03,Rubi04,Santamaria-Holek04}. It was a known fact that the FDT can be violated in many slow relaxation processes~\cite{Parisi97, Vainstein06,Dybiec12}, however, it was a surprise that it could occur in a GLE without  disorder~\cite{cugliandolo94} or  an external field~\cite{Costa03,Dyre00,Vainstein06}.  Finally,  Costa {\it et al.}~\cite{Costa03} have called attention to the fact that if after a long time the fluctuations are not enough to drive the system to equilibrium then the fluctuation-dissipation theorem is violated. For the dissipation relation to be fulfilled, the MC condition, Equation (\ref{MC}), must be valid.

\section{Beyond the basics... and more basic}

   In the last sections we have discussed some basic results for anomalous diffusion under the point of view of the formalism of the generalized Langevin equation which yield two  main features: simplicity and exact results. Obviously, this approach does not exhaust the subject, since diffusion is a basic phenomenon in physics it is a starting point to many different formalisms  which we shall briefly discuss.

\subsection{Fractional Fokker-Planck equation}

We have seen that, in principle, all kinds of anomalous diffusion can  be described by  the GLE formalism, which is itself well established from the Mori method. Since normal  diffusion can be studied both from Langevin equations and from Fokker-Planck equations, we would expect to obtain a generalized Fokker-Planck formalism for  anomalous diffusion. Indeed,  fractal formulations of the Fokker-Planck equation  (FFPE) have been widely used in the literature in the last  decades, in which the evolution of the probability distribution function $P(x,t)$ reads \cite{Metzler99,Metzler00,Metzler04}  
\begin{equation}
\label{FFPE}
\frac{\partial^\alpha P(x,t)}{\partial t^\alpha}=\left[ \frac{1}{m\gamma_\alpha}\frac{\partial U(x)}{\partial x}+D_\alpha \frac{\partial^2}{\partial x^2}\right]P(x,t),
\end{equation}
where on the left-hand-side a fractional Rieman-Liouville time-derivative is defined as \cite{oldham74}
\begin{equation}
\frac{\partial^\alpha P(x,t)}{\partial t^\alpha}=\frac{1}{\Gamma(1-\alpha)}\frac{\partial}{\partial t}\int_0^t\frac{P(x,s)}{(t-s)^\alpha}ds,
\end{equation} 
with $0 < \alpha < 1$. Note that the definition of fractional derivatives is not unique \cite{oldham74,kilbas06}, with a  variety of
possibilities, physically (almost) totally unexplored.
We notice here that the nonlocal character of the fractional Fokker-Planck equation is similar to the memory kernel in the GLE.   On the right hand side, $\gamma_\alpha$ is a generalized friction, $U(x)$ is the potential, while $D_\alpha$  is generalized  Einstein-Stokes relation
\begin{equation}
D_\alpha= \frac{k_BT}{m\gamma_\alpha}.
\end{equation}The major result from  Equation (\ref{FFPE}) for a force-free diffusion  is the asymptotic solution
\begin{equation}
\lim_{t \rightarrow \infty}\langle x^2(t) \rangle = \frac{2D_\alpha}{\Gamma(1+\alpha)} t^\alpha.
\end{equation}
Again, for $\alpha \neq 1$, we reach an anomalous diffusion regime~\cite{Shlesinger93,Metzler00,Metzler04,Klafter96}.

 \subsection{Interface growth }
 
\begin{figure*}[!ht]
\centering
\includegraphics[width=1\textwidth]{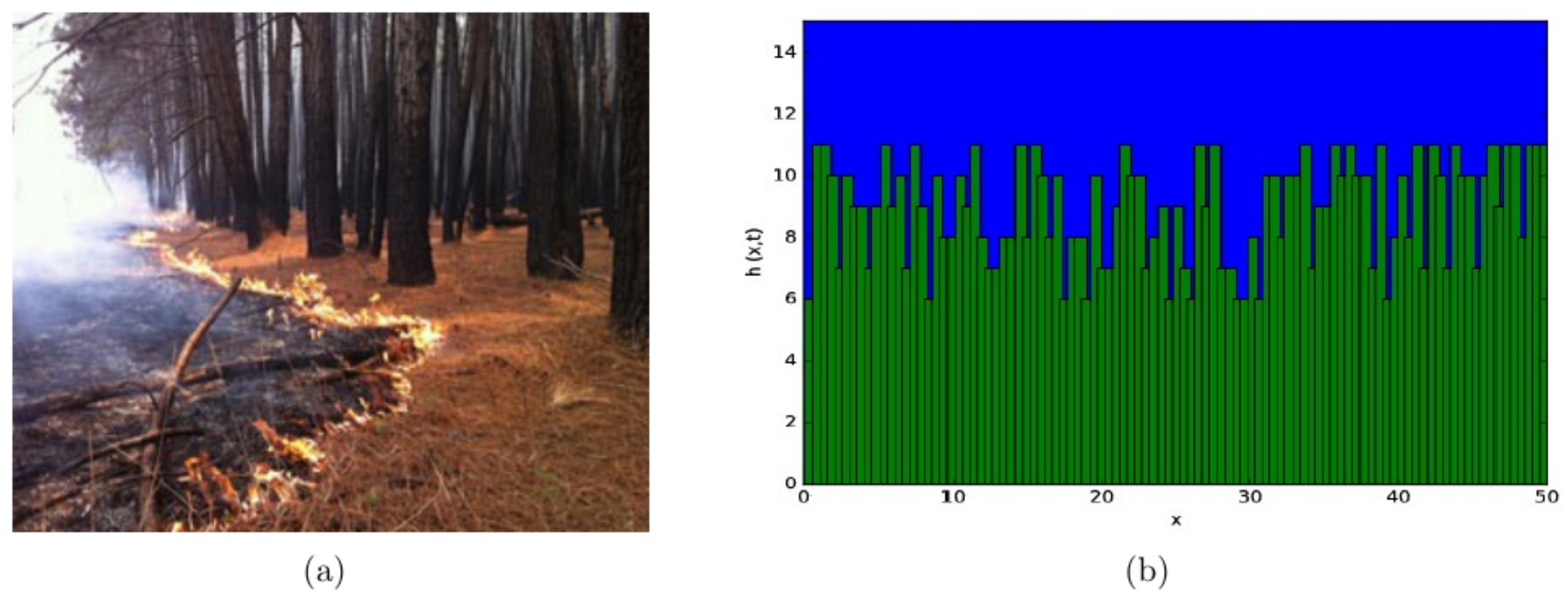}
\caption{(Color online) (a) Fire propagation  in the  Amazon rainforest. Source:
http://g1.globo.com/am/amazonas/noticia/2013/01/florestas-no-sul-do-am-estao-mais-vulneraveis-incendio-aponta-estudo.html;  (b) Schematic representation on a microscopic scale of the height evolution $h(\vec{x},t)$, defining the interface between  two media, as function  of the position $\vec{x}$  for a given time $t$.   }
\label{fig:h}
\end{figure*}

 Models for interface growth generally consider the random deposition of particles that diffuse to a surface and,  as such, have been studied with Langevin equations and modified diffusion equations.  Since diffusion with subsequent deposition is ubiquitous,  rough surfaces at the interface of two media are very common in nature ~\cite{Barabasi95,Edwards82,Kardar86,Hansen00,Cordeiro01,Schmittbuhl06} and  the description of interface evolution is a very interesting problem in statistical physics.  The major objective here  is to  describe the temporal evolution of the height $h(\vec{x},t)$ of the interface between two substrates \cite{Barabasi95,Edwards82},  where $\vec{x}$ is the $d$ dimensional  position vector and $t$ is the time. We outline the evolution of  $h$ in Fig. (\ref{fig:h}). In (a), we show a real  forest fire propagation, a very complex situation. However, we can focus on  the interface between the burnt and unburned regions. In (b), we  provide a snapshot at a fixed time $t$ of the height $h(\vec{x},t)$  for a microscopic growth process in which  the green medium is penetrating the blue one with arbitrary units. 
These types of  dynamics in $d+1$ dimensional  space  are easy to understand,  but not so simple to solve analytically.  Experiments  can be done for $d=1,2,3$;   however, they  present hard theoretical problems for any $d$.

The two main quantities of interest are the average height
\begin{equation}
\label{hmedio}
\langle h(t) \rangle =\frac{1}{V}\int h(\vec{x},t) d\vec{x},
\end{equation}
 where $V$ is the sample volume, and the standard deviation
\begin{equation}
w^2(t)=\langle h^2(t) \rangle - \langle h(t) \rangle^2,
\label{wt}
\end{equation}
 often called the surface width $w(t)$, or roughness.  Not surprisingly, this height fluctuation has a lot of information about the physical processes governing the system.  The evolution of $w(t)$ observed through experiments, computer simulations, and a few analytical  results gives us some general features of growth dynamics.

Starting with a flat surface, $w(0)=0$, the evolution exhibits  four distinct regions: (a) for a very short period $0<t<t_0$,  during which correlations are negligible,  the process is a random deposition $w(t) \sim t^{1/2}$; (b) for $ t_0 < t < t_\times$ we have $ w(t) \sim t^{\beta}$. Here    the  $t_\times$ follows a power law of the form   $t_\times \sim L^{z}$, where $L$ is the size of the sample, $z$ is the dynamic exponent, and $\beta$ the growth exponent; (c) there is a transition region for $ t \sim t_\times$; (d)  finally, for $t \gg t_\times$, the dynamic equilibrium leads to surface width saturation, $w_s$,  which also follows a power law $ w_s  \sim L^{\alpha}$, where $\alpha$ is the roughness exponent.  The crossing of the curves $ w(t) \sim t^{\beta}$ and $w \sim w_s$ yields the universal relation
\begin{equation}
\label{universal}
z=\frac{\alpha}{\beta}.
\end{equation} 
It should be  pointed out  that these exponents are not related with those of the previous section.

To obtain $w(t)$, we need to know $h(\vec{x},t)$ and there are two major theoretical ways to attack this problem: 1.  Continuous  growth equations; 2.   Discrete growth models.   Fig. (\ref{fig:h}a) is an example of the first, while  Fig. (\ref{fig:h}b) is an example of the second.

\subsubsection{Equation of motion and symmetries}
The growth, in general, is due to the number of particles per unit of time $G(\vec{x},t)$ arriving  on the surface at the position $\vec{x}$ and time $t$. The particle flux is not uniform since the particles are deposited at random positions \cite{Barabasi95}. Therefore, the evolution of $h(\vec{x},t)$ can be described by
\begin{equation}
\label{dht}
\frac{\partial h(\vec{x},t)}{\partial t}=G=F+\chi(\vec{x},t),
\end{equation}
where the first term $F$ is the average number of particles arriving at site $\vec{x}$. The second term, $\chi(\vec{x},t)$, reflects the random fluctuations and satisfies
\begin{equation}
\langle \chi({\vec{x},t})\rangle =0,\text{ and }
\end{equation}
\begin{equation}
\langle \chi({\vec{x},t})\chi({\vec{x}\,',t'}) \rangle =2D_g\delta(\vec{x}-\vec{x}\,')\delta(t-t'),
\end{equation}
where, $D_g$ measures the degree of growth randomness.
The deterministic flux $F$ must satisfy certain symmetry requirements, such as invariance under translation of position, $\vec{x} \rightarrow \vec{x}+\vec{x_0}$, height, $h \rightarrow h +h_0$ and time  $t \rightarrow  t +t_0$. To satisfy these conditions,  it must depend only on  derivatives, which, again by the symmetries  $\vec{x}  \rightarrow -\vec{x} $  and $h \rightarrow - h $, must be of even order~\cite{Barabasi95,Edwards82}.  Besides this, considering  that symmetrically possible terms such as $ \nabla^{2n} h(\vec{x},t)$, for $n=2,3,\dotsc$, are irrelevant in the long wavelength limit, since they go to zero faster than $ \nabla^{2} h(\vec{x},t)$,      we obtain, at the lowest-order in the derivatives \cite{Edwards82},
\begin{equation}
\label{EWE}
\frac{\partial h(\vec{x},t)}{\partial t}= \nu \nabla^2 h(\vec{x},t)+\chi(\vec{x},t),
\end{equation}
known as Edwards-Wilkinson equation (EW).  Note that it is basically a diffusion equation, as~(\ref{diff}), plus a noise, where   we have a surface tension $\nu$ associated with the Laplacian  smoothening mechanism. The random deposition model with surface relaxation is 
in the same universality class as the EW model~\cite{Horowitz01}, i.e., they share the same exponents.

Because there are, however, a large class of growth phenomena which are not described by the EW equation, new formulations become necessary.   The recently proposed Arcetri models~\cite{Henkel15} allow the study of more rigid interfaces than the EW model and still allow for exact solutions in any dimension $d$. Depending on the initial conditions, both growing interfaces and particle motion on the lattice can be modeled.  Another classical model   was proposed by Kardar {\it et al.}~\cite{Kardar86},  inspired by the stochastic Burgers equation. They  observed that lateral growth could be added to this equation via the nonlinear term of the Burgers equation so that Equation~(\ref{EWE})  becomes

\begin{equation}
\label{KPZ}
\dfrac{\partial h(\vec{x},t)}{\partial t}=\nu \nabla^2 h(\vec{x},t) +\dfrac{\lambda}{2}[\vec{\nabla}h(\vec{x},t)]^2 + \chi(\vec{x},t).
\end{equation}
Since its formulation, the Kardar-Parisi-Zhang equation (KPZ) has been a prime model in the description of growth dynamics.     The nonlinear term  includes a new  constant $\lambda$ associated with the tilt mechanism and breaks down the symmetry   $h \rightarrow - h $. Consequently, the  universality class of KPZ is different from that of EW.
 
 Equation (\ref{KPZ}) is the simplest nonlinear equation that can describe a large number of growth processes \cite{Barabasi95,Kardar86}.
 However, the apparent simplicity of this equation is misleading, since the nonlinear gradient term combined with the noise makes  it one of the  toughest  problems in modern mathematical physics \cite{Hairer13,Sasamoto10}. On the other hand, its complexity is compensated for its generality.  It is connected to a large number of stochastic processes, such as the direct polymer model~\cite{Kardar85},  the weakly asymmetric simple exclusion process \cite{Bertine97}, the totally asymmetric exclusion process \cite{Spitzer70},  direct d-mer diffusion \cite{Odor10}, fire propagation~\cite{Mylles01,Mylles03,Merikoski03}, 
 atomic deposition~\cite{Csahok92},  evolution of bacterial colonies~\cite{Ben-Jacob94,Matsushita90}, turbulent liquid-crystals \cite{Takeuchi11,Takeuchi12,Takeuchi13}, polymer deposition in semiconductors\cite{Almeida17}, and etching 
\cite{Mello01,Reis03,Reis04,Reis05,Oliveira08,Tang10,Xun12,Rodrigues15,Mello15,Alves16,Carrasco16,Carrasco18}.  

This problem in non-equilibrium statistical physics is analogous to the Ising model for equilibrium statistical physics, which is used as a basic model for understanding a large class of phenomena. The search for exact solutions to this equation in the  $d+1$ dimensional space has resulted in important contributions to mathematics; however, to date, they have been obtained only for specific situations and are limited to  $1+1$ dimensions \cite{Kardar86,Sasamoto10} .

\subsubsection{Scaling invariance}
  In this small subsection we limit ourselves to scaling symmetries only. However, in interface growth, the shape of single-time
and two-time responses and height correlation functions can be derived from the assumption of a local time-dependent scale-invariance~\cite{Henkel17}.     For the KPZ equation,   a systematic test of aging scaling was performed, showing the scaling relations  for the two-time spatio-temporal autocorrelator and for the time-integrated response function~\cite{Reis04, Henkel12,Kelling17a}.

 The scaling invariance can be investigated through the transformation \cite{Barabasi95}  $\vec{x} \rightarrow  b \vec{x}$, $h \rightarrow b^\alpha$ and $t \rightarrow t^z$ which yields, for the KPZ equation, (\ref{KPZ}), 
\begin{equation}
\label{nu}
\nu  \rightarrow b^{z-2} \nu,
\end{equation}
\begin{equation}
\label{Dg}
  D_g  \rightarrow b^{z-d-2\alpha} D_g,
\end{equation}
and
\begin{equation}
\label{lambda}
  \lambda  \rightarrow b^{\alpha+z-2} \lambda
\end{equation}Scaling invariance demands that the exponents in Equations (\ref{nu},\ref{Dg}, \ref{lambda}) must be zero. However, a simple inspection  shows that they are inconsistent.
   The renormalization group approach of Kardar {\it el al.}~\cite{Kardar86}, uses the nonlinear term as a perturbation, and their result shows that only   Equation (\ref{lambda}) remains invariant, yielding the famous Galilean invariance
\begin{equation}
 \alpha+z=2.
 \label{GI}
\end{equation} 
 Equations (\ref{nu}) and (\ref{Dg}) are corrected by the renormalization, and the final result yields $\alpha=1/2$, $\beta=1/3 $, and $z=3/2$, for $1+1$ dimensions. Since the KPZ renormalization approach is valid  only for $1+1$ dimensions, questions about the  validity of the Galilean invariance~\cite{Wio10b,Wio17} for $d>1$ and the existence of an upper critical dimension for KPZ \cite{Francesca01,Schwartz12} have been raised.

  For $d>1$,  the numerical simulation of the KPZ equation is not an easy task~\cite{Wio10b,Wio17,lam98,Xu06,HalpinHealy15,Torres17}, and the use of  cellular automata models~\cite{Mello01,Reis03,Reis04,Reis05,Oliveira08,Tang10,Xun12,Rodrigues15,Mello15,Alves16,Carrasco16,Carrasco18,Kelling17,Predota96,Chua05,Buceta14} has become increasingly common for growth simulations. Polynuclear growth (PNG), is a typical example of a discrete model that has received  a lot of attention, and the outstanding works  of Pr\"ahofer and Spohn~\cite{Prahofer00} and Johansson~\cite{Johansson00}  drive the way to the exact solution of the distributions of the heights fluctuations $f(h,t)$ in the KPZ equation for  $1+1$ dimensions \citep{Sasamoto10}. By construction, $h\rightarrow h-\langle h\rangle $, then $f(h,t)$ has zero mean, so its skewness and kurtosis are the most important quantities to observe \citep{Takeuchi13,Oliveira13,Alves13,Almeida14,Halpin-Healy95}.  In addition, Langevin equations for  growth models have been discussed by some authors~\cite{Haselwandter06,Haselwandter08,Silveira12}. Several works have been done in the weakly asymmetric simple exclusion process \cite{Bertine97}, the totally asymmetric exclusion process \cite{Spitzer70,Alcaraz99}, and the direct d-mer diffusion model \cite{Odor10}: for a review see \citep{HalpinHealy15,Halpin-Healy95,Meakin93,Krug97}.   More experimentally measured exponents  for growing interfaces in four universality classes  (KPZ, quenched KPZ, EW and Arcetri)  can be found in~\cite{Henkel15} and references therein.

\subsubsection{Cellular automata  growth models}

\begin{figure*}[!ht]
\centering
\includegraphics[width=1\textwidth]{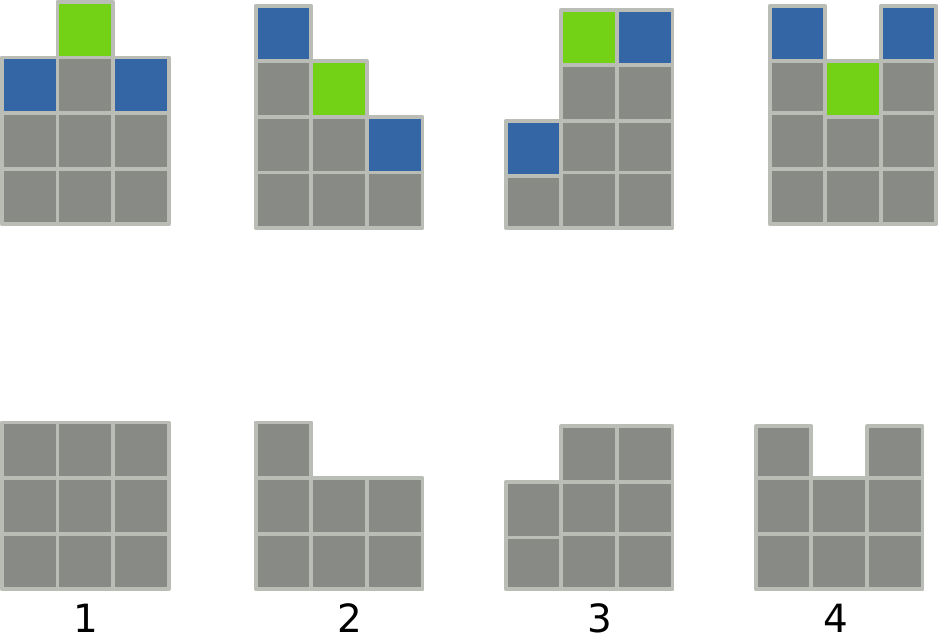}
\caption{ The etching mechanism  in one dimension. The figures are the four situations in which the dynamics of the etching model can be divided. The initial configuration is at the top, where the selected site, in column 2, is green, and its first neighbors, in columns 1 and 3, are shown in dark blue. After the corrosion process the final situation is exhibited at the bottom. The steps are: step 1, site selection, green;  step 2, interaction with nearest neighbors; step 3, decrease of height by one.  From \cite{Alves16},  with permission from Physics Review E.  }
\label{fig:etching}
\end{figure*}

 Cellular automata are described by  simple rules, which  allow us to inquire about relevant properties of a complex dynamical system.  For a given growth model, the first question to be answered is if the model belongs to the same universality class as KPZ. In this context, we expose here the etching model~\cite{Mello01}, which has attracted  considerable attention in recent years~\cite{Reis03,Reis04,Reis05,Oliveira08,Tang10,Xun12,Rodrigues15,Mello15,Alves16,Carrasco16,Carrasco18}.
   These studies suggest a close relation between the etching model and KPZ.  For example, for $1+1$ dimensions Alves {\it et al.}~\cite{Alves16} have proven that $\alpha=1/2$ exactly. Unfortunately, their method does not allow to obtain $\beta$ or $z$. 

The etching model describes	the mechanism of an acid  eroding  a surface. The detailed description of the model can be found in~\cite{Mello01})  For simplicity, let us consider a site $i$ in a hypercube of side $L$ and  volume $V=L^d$ in space $\Psi$ of dimension $d$, and look at one of its  nearest neighbors, $j$. If $h_{j} > h_i$, then it becomes equal to $h_i$.  We then  define  the etching model following the steps:
\begin{enumerate}
\item At time t we randomly choose a site $i\in{V}$.
\item If $h_{j}(t)  >   h_i(t)$,  do $h_{j}(t+\Delta t)=h_i(t)$.
\item Do $h_i(t+\Delta t)=h_i(t)-1.$
\end{enumerate}

In Fig. (\ref{fig:etching}), we show the mechanism of the etching model in one dimension: (a) step 1, we randomly select a site at time $t$, here $i=2$ shown as green in the figure;  step 2, the site $i=2$ interacts with its neighbors $j=1,3$. Then in Fig. (\ref{fig:etching}2) $h_1(t+\Delta t)=h_2(t)$, meaning that it is strongly affected, and   $h_3(t+\Delta t)=h_3(t)$, meaning it is not affected; (c) step 3, $h_2(t+\Delta t)=h_2(t)-1$. We describe here a process of strong interactions between the site and its neighbors. 

Using these rules and  averaging over $N_e$ numerical experiments, it is possible to obtain the exponents $\alpha$, $\beta$, and $z$ \cite{Mello01,Reis03,Reis04,Reis05,Oliveira08,Tang10,Xun12,Rodrigues15,Mello15,Alves16,Carrasco16,Carrasco18}. Finally,  the upper critical limit for the etching model (KPZ) was  recently discussed by Rodrigues {\it et al.}~\cite{Rodrigues15}, where it was shown numerically that there is no upper critical dimension for the etching model up to $6+1$ dimensions.  Thus, we have established  a lower limit for the KPZ critical dimension, i.e., if it exists then $d_c>6$, in agreement with other authors \cite{Odor10}.  They have shown that the Galilean invariance (\ref{GI}) remains valid for $d \leq 6$ as well. These problems, however, still lack an exact solution.

\subsection{Reaction-diffusion processes }

 One could not complete a work on diffusion without a  brief discussion of reaction-diffusion processes \cite{Alcaraz99,ben00,abad02,shapoval18}. Exactly solvable reaction-diffusion models consist largely of single species reactions in one dimension, e.g., variations of the coalescence process, $A + A  \rightarrow  A + S $ \cite{doering89,krebs95,simon95}
and the annihilation process $A + A  \rightarrow  S + S $ \cite{krebs95,simon95}, where $A$ and $S$ denote
occupied and empty sites, respectively. These simple reactions display a wide range of behavior characteristic of non-equilibrium kinetics, such as self organization, pattern formation, and kinetic phase transitions. Interval methods have provided many exact solutions for one-dimensional coalescence and annihilation models. The method of empty
intervals, applicable to coalescence models, requires solution of an infinite hierarchy of differential difference equations for the probabilities $E_n$ of finding $n$ consecutive lattice sites simultaneously empty. For annihilation models, the method of parity intervals similarly requires determination of $G_n$, the probability of $n$ consecutive lattice
sites containing an even number of particles \cite{abad02}. In the continuous limit, such models can be described exactly in terms of the probability $E(t,x)$ of finding an empty interval of size $x$ at time $t$. Under fairly weak conditions, the equation of motion for $E(t,x)$ is a diffusion equation, albeit with the unusual boundary condition $E(t,0)=1$. In several cases, the exact and fluctuation-dominated behaviour in $d=1$ has been seen experimentally, since the 1990s. This is one of the rare cases where theoretical statistical mechanics can be compared with experiments and also shows that simple mean-field schemes are not enough. For an introduction see \cite{ben00}, and for up to date references see \cite{shapoval18}.

Another important new development in diffusion concerns stochastic resets, as they were introduced by  Evans and S.N. Majumdar (see \cite{evans11,evans14,durang14} and references therein).
 This leads to important modifications of the stationary state and raises the question how to consider the status of the general theorem described in the present review.

%%%%%%%%%%%%%%

\section{Conclusions}

In this short review we address diffusion, as a modern and important topic of statistical physics, with broad applications \cite{Metzler00,Morgado02,Metzler04,Costa03, Lapas08,Ciesla01,Holek09}.  We emphasize the study of systems with memory, for which a generalized Langevin equation applies and describe how the diffusion exponent is obtained from the memory \cite{Morgado02,Ferreira12}. We highlight the properties of response functions \cite{Lapas07, Vainstein06} for the processes of anomalous relaxation \cite{Vainstein06,Lapas15}, Gaussianization and ergodicity \cite{Lapas08,Lapas07}. Moreover, we have established a hierarchy: the mixing condition is stronger than the ergodic hypothesis, which is itself stronger than the fluctuation-dissipation theorem. It is important to call attention again to Costa~\cite{Costa03}, where it was observed that in the ballistic diffusion the fluctuation-dissipation theorem fails. Indeed, for the ballistic motion which, on average, behaves as a Newtonian particle with constant  velocity, the fluctuations are not enough to bring the system to  equilibrium, i.e., the dissipation, which for large times decays as $1/t$, does not balance the fluctuations.  This work points out that the violation of the mixing condition breaks down ergodicity, as required by the Khinchin theorem and the fluctuation dissipation theorem.

In conclusion, since the mechanism of diffusion is present in most nonequilibrium processes, diffusion is an exhaustive subject, and we have only called attention to some of its  aspects. Consequently,  we apologize to the authors of important works not mentioned here , such as  aging \cite{cugliandolo94,Reis04,Henkel12,Kelling17a,hodge95} for example.

\section*{Conflict of Interest Statement}

The authors declare that the research was conducted in the absence of any commercial or financial relationships that could be construed as a potential conflict of interest.

\section*{Author Contributions}
FAO suggested the work, and wrote some sections; RMSF was responsible for the section on non-Markovian processes and selected the figures; LCL wrote the ergodicity and decay towards equilibrium sections; MHV collaborated on the section on the generalized Langevin equation and reedited the text.

\section*{Acknowledgments}
This study was financed in part by the Coordenação de Aperfeiçoamento de Pessoal de Nível Superior - Brasil (CAPES) - Finance Code 001, CNPq and FAPDF.  
 MHV was supported by a Senior fellowship (88881.119772/2016-01) from Coordenação de Aperfeiçoamento de Pessoal de Nível Superior (CAPES) at MIT. MHV would like to thank  Prof. Jeff Gore for the hospitality at Physics of Living Systems, MIT.

% Sciences articles please include page numbers in the in-text citations
\bibliographystyle{frontiersinHLTH_FPHY} % for cHealth, Physics and Mathematics articles
\bibliography{referencias}

\begin{thebibliography}{197}
\expandafter\ifx\csname natexlab\endcsname\relax\def\natexlab#1{#1}\fi
\expandafter\ifx\csname urlstyle\endcsname\relax
  \expandafter\ifx\csname doi\endcsname\relax
  \def\doi#1{doi:\discretionary{}{}{}#1}\fi \else
  \expandafter\ifx\csname doi\endcsname\relax
  \def\doi{doi:\discretionary{}{}{}\begingroup \urlstyle{rm}\Url}\fi \fi
\expandafter\ifx\csname selectlanguage\endcsname\relax
  \def\selectlanguage#1{}\fi

\bibitem[{Vainstein et~al.(2006{\natexlab{a}})Vainstein, Costa, and
  Oliveira}]{Vainstein06}
Vainstein MH, Costa IVL, Oliveira FA.
\newblock Mixing, ergodicity and the fluctuation-dissipation theorem in complex
  systems.
\newblock Miguel MC, Rub\'{\i} M, editors, {\em Jamming, Yielding, and
  Irreversible Deformation in Condensed Matter\/} (Springer Berlin /
  Heidelberg) (2006{\natexlab{a}}), {\em Lecture Notes in Phys.\/}, vol. 688,
  159--188.
\newblock \doi{10.1007/11581000_10}.

\bibitem[{Shlesinger et~al.(1993)Shlesinger, Zaslavsky, and
  Klafter}]{Shlesinger93}
Shlesinger MF, Zaslavsky GM, Klafter J.
\newblock Strange kinetics.
\newblock {\em Nature\/} {\bf 363} (1993) 31.
\newblock \doi{10.1038/363031a0}.

\bibitem[{Metzler et~al.(1999)Metzler, Barkai, and Klafter}]{Metzler99}
Metzler R, Barkai E, Klafter J.
\newblock Anomalous diffusion and relaxation close to thermal equilibrium: A
  fractional {F}okker-{P}lanck equation approach.
\newblock {\em Phys. Rev. Lett.\/} {\bf 82} (1999) 3563.
\newblock \doi{10.1103/PhysRevLett.82.3563}.

\bibitem[{Metzler and Klafter(2000)}]{Metzler00}
Metzler R, Klafter J.
\newblock The random walk's guide to anomalous diffusion: a fractional dynamics
  approach.
\newblock {\em Phys. Rep.\/} {\bf 339} (2000) 1.
\newblock \doi{10.1016/S0370-1573(00)00070-3}.

\bibitem[{Morgado et~al.(2002)Morgado, Oliveira, Batrouni, and
  Hansen}]{Morgado02}
Morgado R, Oliveira FA, Batrouni GG, Hansen A.
\newblock {Relation between Anomalous and Normal Diffusion in Systems with
  Memory}.
\newblock {\em Phys. Rev. Lett.\/} {\bf 89} (2002) 100601.
\newblock \doi{10.1103/PhysRevLett.89.100601}.

\bibitem[{Metzler and Klafter(2004)}]{Metzler04}
Metzler R, Klafter J.
\newblock The restaurant at the end of the random walk: recent developments in
  the description of anomalous transport by fractional dynamics.
\newblock {\em J. Phys. A: Math. Gen.\/} {\bf 37} (2004) 161.
\newblock \doi{10.1088/0305-4470/37/31/R01}.

\bibitem[{Sancho et~al.(2004)Sancho, Lacasta, Lindenberg, Sokolov, and
  Romero}]{Sancho04}
Sancho JM, Lacasta AM, Lindenberg K, Sokolov IM, Romero AH.
\newblock Diffusion on a {S}olid {S}urface: {A}nomalous is {N}ormal.
\newblock {\em Phys. Rev. Lett.\/} {\bf 92} (2004) 250601.
\newblock \doi{10.1103/PhysRevLett.92.250601}.

\bibitem[{Costa et~al.(2003)Costa, Morgado, Lima, and Oliveira}]{Costa03}
Costa IVL, Morgado R, Lima MVBT, Oliveira FA.
\newblock {The Fluctuation-Dissipation Theorem fails for fast superdiffusion}.
\newblock {\em Europhys. Lett.\/} {\bf 63} (2003) 173.
\newblock \doi{10.1209/epl/i2003-00514-3}.

\bibitem[{Lapas et~al.(2008)Lapas, Morgado, Vainstein, Rub\'{\i}, and
  Oliveira}]{Lapas08}
Lapas LC, Morgado R, Vainstein MH, Rub\'{\i} JM, Oliveira FA.
\newblock Khinchin theorem and anomalous diffusion.
\newblock {\em Phys. Rev. Lett.\/} {\bf 101} (2008) 230602.
\newblock \doi{10.1103/PhysRevLett.101.230602}.

\bibitem[{Weron and Magdziarz(2010)}]{Weron10}
Weron A, Magdziarz M.
\newblock Generalization of the {K}hinchin theorem to {L}\'evy flights.
\newblock {\em Phys. Rev. Lett.\/} {\bf 105} (2010) 260603.
\newblock \doi{10.1103/PhysRevLett.105.260603}.

\bibitem[{Thiel et~al.(2013)Thiel, Flegel, and Sokolov}]{Thiel13}
Thiel F, Flegel F, Sokolov IM.
\newblock Disentangling sources of anomalous diffusion.
\newblock {\em Phys. Rev. Lett.\/} {\bf 111} (2013) 010601.
\newblock \doi{10.1103/PhysRevLett.111.010601}.

\bibitem[{McKinley and Nguyen(2018)}]{Mckinley18}
McKinley S, Nguyen H.
\newblock Anomalous diffusion and the generalized langevin equation.
\newblock {\em SIAM Journal on Mathematical Analysis\/} {\bf 50} (2018)
  5119--5160.
\newblock \doi{10.1137/17M115517X}.

\bibitem[{Flekk\o{}y(2017)}]{Flekkoy17}
Flekk\o{}y EG.
\newblock Minimal model for anomalous diffusion.
\newblock {\em Phys. Rev. E\/} {\bf 95} (2017) 012139.
\newblock \doi{10.1103/PhysRevE.95.012139}.

\bibitem[{Barab\'asi and Stanley(1995)}]{Barabasi95}
Barab\'asi AL, Stanley HE.
\newblock {\em {Fractal Concepts in Surface Growth}\/} (Cambridge: Cambridge
  University Press) (1995).

\bibitem[{Murray(2002)}]{Murray2002}
Murray JD.
\newblock {\em Mathematical Biology I. An Introduction\/}, {\em
  Interdisciplinary Applied Mathematics\/}, vol.~17 (New York: Springer), 3
  edn. (2002).
\newblock \doi{10.1007/b98868}.

\bibitem[{Brown(1828{\natexlab{a}})}]{Brown28}
Brown R.
\newblock {A Brief Account of Microscopical Observations Made in the Months on
  June, July, and August, 1827, on the Particles Contained in the Pollen of
  Plants; and on the General Existence of Active Molecules in Organic and
  Inorganic Bodies.}
\newblock {\em Phil. Mag.\/} {\bf 4} (1828{\natexlab{a}}) 161.
\newblock \doi{10.1080/14786442808674769}.

\bibitem[{Brown(1828{\natexlab{b}})}]{Brown28a}
Brown R.
\newblock {Mikroskopische Beobachtungen über die im Pollen der Pflanzen
  enthaltenen Partikeln, und über das allgemeine Vorkommen activer Molecüle
  in organischen und unorganischen Körpern}.
\newblock {\em Annalen der Physik\/} {\bf 90} (1828{\natexlab{b}}) 294--313.
\newblock \doi{10.1002/andp.18280901016}.

\bibitem[{Einstein(1905)}]{Einstein1905}
Einstein A.
\newblock Über die von der molekularkinetischen {T}heorie der {W}ärme
  geforderte {B}ewegung von in ruhenden {F}lüssigkeiten suspendierten
  {T}eilchen.
\newblock {\em Annalen der Physik\/} {\bf 322} (1905) 549--560.
\newblock \doi{10.1002/andp.19053220806}.

\bibitem[{Einstein(1956)}]{Einstein56}
Einstein A.
\newblock {\em {I}nvestigations on the theory of the {B}rownian {M}ovement\/}
  (New York: Dover) (1956).

\bibitem[{Dyre and Schr\o{}der(2000)}]{Dyre00}
Dyre JC, Schr\o{}der TB.
\newblock Universality of {AC} conduction in disordered solids.
\newblock {\em Rev. Mod. Phys.\/} {\bf 72} (2000) 873--892.
\newblock \doi{10.1103/RevModPhys.72.873}.

\bibitem[{Oliveira et~al.(2005)Oliveira, Morgado, Hansen, and
  Rub\'{\i}}]{Oliveira05}
Oliveira FA, Morgado R, Hansen A, Rub\'{\i} JM.
\newblock Superdiffusive conduction: {AC} conductivity with correlated noise.
\newblock {\em Physica A\/} {\bf 357} (2005) 115--121.
\newblock \doi{10.1016/j.physa.2005.05.056}.

\bibitem[{Gudowska-Nowak et~al.(2017)Gudowska-Nowak, Lindenberg, and
  Metzler}]{Nowak17}
Gudowska-Nowak E, Lindenberg K, Metzler R.
\newblock Preface: {M}arian {S}moluchowski’s 1916 paper—a century of
  inspiration.
\newblock {\em J. Phys. A\/} {\bf 50} (2017) 380301.
\newblock \doi{10.1088/1751-8121/aa8529}.

\bibitem[{Risken(1989)}]{Risken89}
Risken H.
\newblock {\em The {F}okker-{P}lanck equation\/} (Berlin: Springer-Verlag)
  (1989).

\bibitem[{Salinas(2001)}]{Salinas01}
Salinas S.
\newblock {\em Introduction to Statistical Physics\/} (New York: Springer)
  (2001).

\bibitem[{Gadomski et~al.(2018)Gadomski, Kruszewska, Be{\l}dowski, Lent, and
  Ausloos}]{Gadomski18}
Gadomski A, Kruszewska N, Be{\l}dowski P, Lent B, Ausloos M.
\newblock A tribute to {M}arian {S}moluchowski's legacy on soft grains assembly
  and hydrogel formation.
\newblock {\em Acta Phys. Polon. B\/} {\bf 49} (2018) 993--1005.

\bibitem[{Smo(2017)}]{Smoluchowski17}
{On the Uniformity of Laws of Nature. XXX Marian Smoluchowski Symposium}
  (Krakow, Poland) (2017).
\newblock 100$^{th}$ anniversary of {S}moluchowski's death. Available online
  at: http://www.smoluchowski.if.uj.edu.pl/smoluchowski-2017.

\bibitem[{Langevin(1908)}]{Langevin08}
Langevin P.
\newblock Sur la théorie du mouvement {B}rownien.
\newblock {\em C. R. Acad. Sci. (Paris)\/} {\bf 146} (1908) 530.

\bibitem[{Reichl(1998)}]{Reichl98}
Reichl LE.
\newblock {\em A modern course in statistical physics\/} (New York:
  Wiley-Interscience) (1998).

\bibitem[{Toussaint et~al.(2004)Toussaint, Helgesen, and
  Flekk{\o}y}]{Toussaint04}
Toussaint R, Helgesen G, Flekk{\o}y EG.
\newblock {Dynamic Roughening and Fluctuations of Dipolar Chains}.
\newblock {\em Phys. Rev. Lett.\/} {\bf 93} (2004) 108304.
\newblock \doi{10.1103/PhysRevLett.93.108304}.

\bibitem[{Oliveira and Taylor(1994)}]{Oliveira94}
Oliveira FA, Taylor PL.
\newblock Breaking in polymer chains. {II}. {T}he {L}ennard‐{J}ones chain.
\newblock {\em J. Chem. Phys.\/} {\bf 101} (1994) 10118--10125.
\newblock \doi{10.1063/1.468000}.

\bibitem[{Oliveira and Gonzalez(1996)}]{Oliveira96}
Oliveira FA, Gonzalez JA.
\newblock Bond-stability criterion in chain dynamics.
\newblock {\em Phys. Rev. B\/} {\bf 54} (1996) 3954--3958.
\newblock \doi{10.1103/PhysRevB.54.3954}.

\bibitem[{Oliveira(1998)}]{Oliveira98a}
Oliveira FA.
\newblock Transition-state analysis for fracture nucleation in polymers: {T}he
  {L}ennard-{J}ones chain.
\newblock {\em Phys. Rev. B\/} {\bf 57} (1998) 10576--10582.
\newblock \doi{10.1103/PhysRevB.57.10576}.

\bibitem[{Maroja et~al.(2001)Maroja, Oliveira, Cie\ifmmode~\acute{s}\else
  \'{s}\fi{}la, and Longa}]{Maroja01}
Maroja AM, Oliveira FA, Cie\ifmmode~\acute{s}\else \'{s}\fi{}la M, Longa L.
\newblock Polymer fragmentation in extensional flow.
\newblock {\em Phys. Rev. E\/} {\bf 63} (2001) 061801.
\newblock \doi{10.1103/PhysRevE.63.061801}.

\bibitem[{Dias et~al.(2005)Dias, Dub\'e, Oliveira, and Grant}]{Dias05}
Dias CL, Dub\'e M, Oliveira FA, Grant M.
\newblock Scaling in force spectroscopy of macromolecules.
\newblock {\em Phys. Rev. E\/} {\bf 72} (2005) 011918.
\newblock \doi{10.1103/PhysRevE.72.011918}.

\bibitem[{Rahman et~al.(1962)Rahman, Singwi, and Sj\"olander}]{Rahman62}
Rahman A, Singwi KS, Sj\"olander A.
\newblock Stochastic model of a liquid and cold neutron scattering. {II}.
\newblock {\em Phys. Rev.\/} {\bf 126} (1962) 997--1004.
\newblock \doi{10.1103/PhysRev.126.997}.

\bibitem[{Yulmetyev et~al.(2003)Yulmetyev, Mokshin, and H\"anggi}]{Yulmetyev03}
Yulmetyev RM, Mokshin AV, H\"anggi P.
\newblock Diffusion time-scale invariance, randomization processes, and memory
  effects in {L}ennard-{J}ones liquids.
\newblock {\em Phys. Rev. E\/} {\bf 68} (2003) 051201.
\newblock \doi{10.1103/PhysRevE.68.051201}.

\bibitem[{Bao(2003)}]{Bao03a}
Bao JD.
\newblock Transport in a flashing ratchet in the presence of anomalous
  diffusion.
\newblock {\em Phys. Lett. A\/} {\bf 314} (2003) 203 -- 208.
\newblock \doi{10.1016/S0375-9601(03)00910-1}.

\bibitem[{Bao et~al.(2006)Bao, Zhuo, Oliveira, and H\"{a}nggi}]{Bao06}
Bao JD, Zhuo YZ, Oliveira FA, H\"{a}nggi P.
\newblock Intermediate dynamics between {N}ewton and {L}angevin.
\newblock {\em Phys. Rev. E\/} {\bf 74} (2006) 061111.
\newblock \doi{10.1103/PhysRevE.74.061111}.

\bibitem[{Longa et~al.(1996)Longa, Curado, and Oliveira}]{Longa96}
Longa L, Curado EMF, Oliveira FA.
\newblock Roundoff-induced coalescence of chaotic trajectories.
\newblock {\em Phys. Rev. E\/} {\bf 54} (1996) R2201.

\bibitem[{Cie\'sla et~al.(2001)Cie\'sla, Dias, Longa, and Oliveira}]{Ciesla01}
Cie\'sla M, Dias SP, Longa L, Oliveira FA.
\newblock Synchronization induced by {L}angevin dynamics.
\newblock {\em Phys. Rev. E\/} {\bf 63} (2001) 065202.
\newblock \doi{10.1103/PhysRevE.63.065202}.

\bibitem[{Huang(1987)}]{Huang87}
Huang K.
\newblock {\em Statistical {M}echanics\/} (New york: John Wiley \& Sons)
  (1987).

\bibitem[{Santamaría-Holek et~al.(2009)Santamaría-Holek, Vainstein,
  Rub\'{\i}, and Oliveira}]{Holek09}
Santamaría-Holek I, Vainstein MH, Rub\'{\i} JM, Oliveira FA.
\newblock Protein motors induced enhanced diffusion in intracellular transport.
\newblock {\em Physica A\/} {\bf 388} (2009) 1515 -- 1520.
\newblock \doi{10.1016/j.physa.2009.01.013}.

\bibitem[{Palmieri et~al.(2015)Palmieri, Bresler, Wirtz, and
  Grant}]{Palmieri15}
Palmieri B, Bresler Y, Wirtz D, Grant M.
\newblock Multiple scale model for cell migration in monolayers: Elastic
  mismatch between cells enhances motility.
\newblock {\em Sci. Rep.\/} {\bf 5} (2015) 11745.
\newblock \doi{10.1038/srep11745}.

\bibitem[{Lomholt et~al.(2005)Lomholt, Ambj\"ornsson, and Metzler}]{Lomholt05}
Lomholt MA, Ambj\"ornsson T, Metzler R.
\newblock Optimal target search on a fast-folding polymer chain with volume
  exchange.
\newblock {\em Phys. Rev. Lett.\/} {\bf 95} (2005) 260603.
\newblock \doi{10.1103/PhysRevLett.95.260603}.

\bibitem[{Durang et~al.(2015)Durang, Kwon, and Park}]{Durang15}
Durang X, Kwon C, Park H.
\newblock Overdamped limit and inverse-friction expansion for {B}rownian motion
  in an inhomogeneous medium.
\newblock {\em Phys. Rev. E\/} {\bf 91} (2015) 062118.
\newblock \doi{10.1103/PhysRevE.91.062118}.

\bibitem[{Morgado et~al.(2004)Morgado, Costa, and Oliveira}]{Morgado04}
Morgado R, Costa IVL, Oliveira FA.
\newblock Normal and anomalous diffusion: Ergodicity and
  fluctuation-dissipation theorem.
\newblock {\em Acta Phys. Polon. B\/} {\bf 35} (2004) 1359.

\bibitem[{{Vaytet, N.} et~al.(2018){Vaytet, N.}, {Commer\c{c}on, B.}, {Masson,
  J.}, {Gonz\'alez, M.}, and {Chabrier, G.}}]{Vaytet18}
{Vaytet, N}, {Commer\c{c}on, B}, {Masson, J}, {Gonz\'alez, M}, {Chabrier, G}.
\newblock Protostellar birth with ambipolar and ohmic diffusion.
\newblock {\em A\&A\/} {\bf 615} (2018) A5.
\newblock \doi{10.1051/0004-6361/201732075}.

\bibitem[{Mason and Weitz(1995)}]{Mason95}
Mason TG, Weitz D.
\newblock Optical measurements of frequency-dependent linear viscoelastic
  moduli of complex fluids.
\newblock {\em Phys. Rev. Lett.\/} {\bf 74} (1995) 1250.

\bibitem[{Grmela and {\"O}ttinger(1997)}]{Grmela97}
Grmela M, {\"O}ttinger HC.
\newblock Dynamics and thermodynamics of complex fluids. {I. D}evelopment of a
  general formalism.
\newblock {\em Phys. Rev. E\/} {\bf 56} (1997) 6620.

\bibitem[{Bakk et~al.(2002)Bakk, Fossum, da~Silva, Adland, Mikkelsen, and
  Elgsaeter}]{Bakk02}
Bakk A, Fossum JO, da~Silva GJ, Adland HM, Mikkelsen A, Elgsaeter A.
\newblock Viscosity and transient electric birefringence study of clay
  colloidal aggregation.
\newblock {\em Phys. Rev. E\/} {\bf 65} (2002) 021407.

\bibitem[{Sehnem et~al.(2014)Sehnem, Aquino, Campos, Tourinho, Depeyrot, and
  Neto}]{Sehnem2014}
Sehnem AL, Aquino R, Campos AFC, Tourinho FA, Depeyrot J, Neto AMF.
\newblock Thermodiffusion in positively charged magnetic colloids: Influence of
  the particle diameter.
\newblock {\em Phys. Rev. E\/} {\bf 89} (2014) 032308.

\bibitem[{Sehnem et~al.(2015)Sehnem, Neto, Aquino, Campos, Tourinho, and
  Depeyrot}]{Sehnem15}
Sehnem AL, Neto AMF, Aquino R, Campos AFC, Tourinho FA, Depeyrot J.
\newblock Temperature dependence of the {S}oret coefficient of ionic colloids.
\newblock {\em Phys. Rev. E\/} {\bf 92} (2015) 042311.

\bibitem[{Cabreira~Gomes et~al.(2018)Cabreira~Gomes, Ferreira~da Silva,
  Kouyaté, Demouchy, Mériguet, Aquino et~al.}]{cabreira18}
Cabreira~Gomes R, Ferreira~da Silva A, Kouyaté M, Demouchy G, Mériguet G,
  Aquino R, et~al.
\newblock Thermodiffusion of repulsive charged nanoparticles – the interplay
  between single-particle and thermoelectric contributions.
\newblock {\em Phys. Chem. Chem. Phys.\/} {\bf 20} (2018) 16402--16413.
\newblock \doi{10.1039/C8CP02558D}.

\bibitem[{{de Brito} et~al.(1995){de Brito}, {da Silva}, and
  Nazareno}]{deBrito95}
{de Brito} PE, {da Silva} CAA, Nazareno HN.
\newblock {Field-induced localization in Fibonacci and Thue-Morse lattices}.
\newblock {\em Phys. Rev. B\/} {\bf 51} (1995) 6096--6099.

\bibitem[{Monte et~al.(2000)Monte, {da Silva}, Cruz, Morais, Chaves, and
  Cox}]{Monte00}
Monte AFG, {da Silva} SW, Cruz JMR, Morais PC, Chaves AS, Cox HM.
\newblock Symmetric and asymmetric fractal diffusion of electron-hole plasmas
  in semiconductor quantum wells.
\newblock {\em Phys. Lett. A\/} {\bf 268} (2000) 430--435.

\bibitem[{Monte et~al.(2002)Monte, {da Silva}, Cruz, Morais, and
  Chaves}]{Monte02}
Monte AFG, {da Silva} SW, Cruz JMR, Morais PC, Chaves AS.
\newblock {Experimental evidence of asymmetric carrier transport in InGaAs
  quantum wells and wires grown on tilted InP substrates}.
\newblock {\em Appl. Phys. Lett.\/} {\bf 81} (2002) 2460--2462.

\bibitem[{Kumakura et~al.(2005)Kumakura, Makimoto, Kobayashi, Hashizume, Fukui,
  and Hasegawa}]{Kumakura05}
Kumakura K, Makimoto T, Kobayashi N, Hashizume T, Fukui T, Hasegawa H.
\newblock Minority carrier diffusion length in {GaN}: Dislocation density and
  doping concentration dependence.
\newblock {\em Appl. Phys. Lett.\/} {\bf 86} (2005) 052105.

\bibitem[{Borges et~al.(2006)Borges, {da Silva}, Morais, and Monte}]{Borges06}
Borges JB, {da Silva} SW, Morais PC, Monte AFG.
\newblock Optical signatures of asymmetric fractal diffusion of electron-hole
  plasma in semiconductor quantum wells.
\newblock {\em Appl. Phys. Lett.\/} {\bf 89} (2006) 142103.

\bibitem[{Gudowska-Nowak et~al.(2005)Gudowska-Nowak, Bochenek, Jurlewicz, and
  Weron}]{Novak05}
Gudowska-Nowak E, Bochenek K, Jurlewicz A, Weron K.
\newblock Hopping models of charge transfer in a complex environment: Coupled
  memory continuous-time random walk approach.
\newblock {\em Phys. Rev. E\/} {\bf 72} (2005) 061101.
\newblock \doi{10.1103/PhysRevE.72.061101}.

\bibitem[{Filipovitch et~al.(2016)Filipovitch, Hill, Longjas, and
  Voller}]{Filipovitch16}
Filipovitch N, Hill K, Longjas A, Voller V.
\newblock Infiltration experiments demonstrate an explicit connection between
  heterogeneity and anomalous diffusion behavior.
\newblock {\em Water Resour. Res.\/} {\bf 52} (2016) 5167--5178.

\bibitem[{{Aar\~ao Reis}(2016)}]{Reis16}
{Aar\~ao Reis} FDA.
\newblock Scaling relations in the diffusive infiltration in fractals.
\newblock {\em Phys. Rev. E\/} {\bf 94} (2016) 052124.

\bibitem[{{Aar{\~a}o Reis} et~al.(2018){Aar{\~a}o Reis}, Bolster, and
  Voller}]{Reis18}
{Aar{\~a}o Reis} FDA, Bolster D, Voller VR.
\newblock Anomalous behaviors during infiltration into heterogeneous porous
  media.
\newblock {\em Adv. Water Resour.\/} {\bf 113} (2018) 180--188.

\bibitem[{Gomes~Filho et~al.(2016)Gomes~Filho, Oliveira, and
  Barbosa}]{GomesFilho16}
Gomes~Filho MS, Oliveira FA, Barbosa MAA.
\newblock A statistical mechanical model for drug release: Investigations on
  size and porosity dependence.
\newblock {\em Physica A\/} {\bf 460} (2016) 29--37.
\newblock \doi{10.1016/j.physa.2016.04.040}.

\bibitem[{Ignacio et~al.(2017)Ignacio, Chubynsky, and Slater}]{Ignacio17}
Ignacio M, Chubynsky MV, Slater GW.
\newblock Interpreting the {W}eibull fitting parameters for
  diffusion-controlled release data.
\newblock {\em Physica A\/} {\bf 486} (2017) 486--496.

\bibitem[{Gun et~al.(2017)Gun, Kaikai, and Rui}]{Gun17}
Gun L, Kaikai Z, Rui G.
\newblock Simulation on drug molecules permeability of the blood-brain-barrier.
\newblock {\em Am. J. Biol. Life Sci.\/} {\bf 5} (2017) 30.

\bibitem[{Soares et~al.(2017)Soares, Almeida, and {de Oliveira}}]{Soares17}
Soares JMD, Almeida JRGS, {de Oliveira} HP.
\newblock Controlled release of extract of {M}orus nigra from {E}udragit
  {L}-100 electrospun fibers: Toxicity and in vitro release evaluation.
\newblock {\em Curr. Trad. Med.\/} {\bf 3} (2017) 146--154.

\bibitem[{Mandelbrot(1982)}]{Mandelbrot82}
Mandelbrot BB.
\newblock {\em The fractal geometry of nature\/}, vol.~1 (WH freeman New York)
  (1982).

\bibitem[{Stauffer and Stanley(1995)}]{Stauffer95}
Stauffer D, Stanley HE.
\newblock {\em {From Newton to Mandelbrot}\/} (New York: Springer-Verlag)
  (1995).

\bibitem[{Cristea and Steinsky(2014)}]{Cristea14}
Cristea LL, Steinsky B.
\newblock On totally disconnected generalised {S}ierpi{\'n}ski carpets.
\newblock {\em B. Math. Soc. Sci. Math.\/}  (2014) 27--34.

\bibitem[{Balankin(2017)}]{Balankin17}
Balankin AS.
\newblock The topological {H}ausdorff dimension and transport properties of
  {S}ierpi{\'n}ski carpets.
\newblock {\em Phys. Lett. A\/} {\bf 381} (2017) 2801--2808.

\bibitem[{Balankin(2018)}]{Balankin18}
Balankin AS.
\newblock Mapping physical problems on fractals onto boundary value problems
  within continuum framework.
\newblock {\em Phys. Lett. A\/} {\bf 382} (2018) 141--146.

\bibitem[{Barbosa et~al.(2011)Barbosa, Barbosa, and Oliveira}]{Barbosa11}
Barbosa MAA, Barbosa FV, Oliveira FA.
\newblock Thermodynamic and dynamic anomalies in a one-dimensional lattice
  model of liquid water.
\newblock {\em J. Chem. Phys.\/} {\bf 134} (2011) 024511.

\bibitem[{Bertolazzo et~al.(2015)Bertolazzo, Kumar, Chakravarty, and
  Molinero}]{Bertolazzo15}
Bertolazzo AA, Kumar A, Chakravarty C, Molinero V.
\newblock Water-like anomalies and phase behavior of a pair potential that
  stabilizes diamond.
\newblock {\em J. Phys. Chem. B\/} {\bf 120} (2015) 1649--1659.

\bibitem[{da~Silva et~al.(2015)da~Silva, Oliveira, and Barbosa}]{Silva15}
da~Silva FBV, Oliveira FA, Barbosa MAA.
\newblock Residual entropy and waterlike anomalies in the repulsive one
  dimensional lattice gas.
\newblock {\em J. Chem. Phys.\/} {\bf 142} (2015) 144506.

\bibitem[{Bier et~al.(2016)Bier, Lisowski, and Gudowska-Nowak}]{Bier16}
Bier M, Lisowski B, Gudowska-Nowak E.
\newblock Phase transitions and entropies for synchronizing oscillators.
\newblock {\em Phys. Rev. E\/} {\bf 93} (2016) 012143.

\bibitem[{Pinto et~al.(2016)Pinto, Oliveira, and Penna}]{Pinto16}
Pinto PD, Oliveira FA, Penna ALA.
\newblock Thermodynamics aspects of noise-induced phase synchronization.
\newblock {\em Phys. Rev. E\/} {\bf 93} (2016) 052220.
\newblock \doi{10.1103/PhysRevE.93.052220}.

\bibitem[{Pinto et~al.(2017)Pinto, Penna, and Oliveira}]{Pinto17}
Pinto PD, Penna AL, Oliveira FA.
\newblock Critical behavior of noise-induced phase synchronization.
\newblock {\em EPL (Europhysics Letters)\/} {\bf 117} (2017) 50009.

\bibitem[{{P\'erez-Madrid}(2004)}]{Perez-Madrid04a}
{P\'erez-Madrid} A.
\newblock Gibbs entropy and irreversibility.
\newblock {\em Physica A\/} {\bf 339} (2004) 339.
\newblock \doi{10.1016/j.physa.2004.04.106}.

\bibitem[{Rub\'{\i} and Bedeaux(1988)}]{Rubi88}
Rub\'{\i} JM, Bedeaux D.
\newblock Brownian motion in a fluid in elongational flow.
\newblock {\em J. Stat. Phys.\/} {\bf 53} (1988) 125.
\newblock \doi{10.1007/BF01011549}.

\bibitem[{Ku{\'s}mierz et~al.(2018)Ku{\'s}mierz, Dybiec, and
  Gudowska-Nowak}]{kusmierz18}
Ku{\'s}mierz {\L}, Dybiec B, Gudowska-Nowak E.
\newblock Thermodynamics of superdiffusion generated by {L}{\'e}vy--{W}iener
  fluctuating forces.
\newblock {\em Entropy\/} {\bf 20} (2018) 658.

\bibitem[{H\"anggi et~al.(1990)H\"anggi, Talkner, and Borkovec}]{Hanggi90}
H\"anggi P, Talkner P, Borkovec M.
\newblock Reaction-rate theory: fifty years after {K}ramers.
\newblock {\em Rev. Mod. Phys.\/} {\bf 62} (1990) 251--341.
\newblock \doi{10.1103/RevModPhys.62.251}.

\bibitem[{Morgado et~al.(2007)Morgado, Cie\'sla, Longa, and
  Oliveira}]{Morgado07}
Morgado R, Cie\'sla M, Longa L, Oliveira FA.
\newblock Synchronization in the presence of memory.
\newblock {\em Europhys. Lett.\/} {\bf 79} (2007) 10002.
\newblock \doi{10.1209/0295-5075/79/10002}.

\bibitem[{Lapas et~al.(2007)Lapas, Costa, Vainstein, and Oliveira}]{Lapas07}
Lapas LC, Costa IVL, Vainstein MH, Oliveira FA.
\newblock Entropy, non-ergodicity and non-gaussian behaviour in ballistic
  transport.
\newblock {\em Europhys. Lett.\/} {\bf 77} (2007) 37004.
\newblock \doi{10.1209/0295-5075/77/37004}.

\bibitem[{Lee(1983{\natexlab{a}})}]{Lee83}
Lee MH.
\newblock Can the velocity autocorrelation function decay exponentially?
\newblock {\em Phys. Rev. Lett.\/} {\bf 51} (1983{\natexlab{a}}) 1227--1230.
\newblock \doi{10.1103/PhysRevLett.51.1227}.

\bibitem[{Nyquist(1928)}]{Nyquist28}
Nyquist H.
\newblock Thermal agitation of electric charge in conductors.
\newblock {\em Phys. Rev.\/} {\bf 32} (1928) 110--113.
\newblock \doi{10.1103/PhysRev.32.110}.

\bibitem[{Mori(1965)}]{Mori65}
Mori H.
\newblock Transport, {C}ollective {M}otion, and {B}rownian {M}otion.
\newblock {\em Prog. Theor. Phys.\/} {\bf 33} (1965) 423.

\bibitem[{Kubo(1974)}]{Kubo74}
Kubo R.
\newblock Response, relaxation and fluctuation.
\newblock Kirczenow G, Marro J, editors, {\em Transport {P}henomena. Lecture
  {Notes in Phys.}\/} (Berlin, Heidelberg: Springer-Verlag) (1974), vol.~31,
  74--124.
\newblock \doi{10.1007/3-540-06955-0_3}.

\bibitem[{Kubo et~al.(1991)Kubo, Toda, and Hashitsume}]{Kubo91}
Kubo R, Toda M, Hashitsume N.
\newblock {\em Statistical {P}hysics {II}\/} (Berlin: Springer) (1991).

\bibitem[{Kubo(1966)}]{Kubo66}
Kubo R.
\newblock Fluctuation-dissipation theorem.
\newblock {\em Rep. Prog. Phys.\/} {\bf 29} (1966) 255.
\newblock \doi{10.1088/0034-4885/29/1/306}.

\bibitem[{Lee(1982)}]{Lee82}
Lee MH.
\newblock Solutions of the generalized {L}angevin equation by a method of
  recurrence relations.
\newblock {\em Phys. Rev. B\/} {\bf 26} (1982) 2547.

\bibitem[{Lee(1983{\natexlab{b}})}]{Lee83a}
Lee MH.
\newblock Derivation of the generalized {L}angevin equation by a method of
  recurrence relations.
\newblock {\em J. Math. Phys.\/} {\bf 24} (1983{\natexlab{b}}) 2512.

\bibitem[{Lee and Hong(1984)}]{Lee84}
Lee MH, Hong J.
\newblock Transport behavior of dense protons in a slab.
\newblock {\em Phys. Rev. B\/} {\bf 30} (1984) 6756.

\bibitem[{Gluskin(2003)}]{Gluskin03}
Gluskin E.
\newblock Let us teach this generalization of the final-value theorem.
\newblock {\em Eur. J. Phys.\/} {\bf 24} (2003) 591.

\bibitem[{Vainstein et~al.(2005)Vainstein, Morgado, Oliveira, {de Moura}, and
  {Coutinho-Filho}}]{Vainstein05}
Vainstein MH, Morgado R, Oliveira FA, {de Moura} FABF, {Coutinho-Filho} MD.
\newblock Stochastic description of the dynamics of the random-exchange
  {H}eisenberg chain.
\newblock {\em Phys. Lett. A\/} {\bf 339} (2005) 33--38.
\newblock \doi{10.1016/j.physleta.2005.02.059}.

\bibitem[{Vainstein et~al.(2006{\natexlab{b}})Vainstein, Costa, Morgado, and
  Oliveira}]{Vainstein06a}
Vainstein MH, Costa IVL, Morgado R, Oliveira FA.
\newblock Non-exponential relaxation for anomalous diffusion.
\newblock {\em Europhys. Lett.\/} {\bf 73} (2006{\natexlab{b}}) 726--732.
\newblock \doi{10.1209/epl/i2005-10455-9}.

\bibitem[{Ferreira et~al.(2012)Ferreira, Santos, Donato, Andrade, and
  Oliveira}]{Ferreira12}
Ferreira RMS, Santos MVS, Donato CC, Andrade JS, Oliveira FA.
\newblock Analytical results for long-time behavior in anomalous diffusion.
\newblock {\em Phys. Rev. E\/} {\bf 86} (2012) 021121.
\newblock \doi{10.1103/PhysRevE.86.021121}.

\bibitem[{Srokowski(2000)}]{Srokowski00}
Srokowski T.
\newblock {Nonstationarity Induced by Long-Time Noise Correlations in the
  Langevin Equation}.
\newblock {\em Phys. Rev. Lett.\/} {\bf 85} (2000) 2232.
\newblock \doi{10.1103/PhysRevLett.85.2232}.

\bibitem[{Srokowski(2013)}]{Srokowski13}
Srokowski T.
\newblock Fluctuations in multiplicative systems with jumps.
\newblock {\em Phys. Rev. E\/} {\bf 87} (2013) 032104.

\bibitem[{Kadanoff et~al.(1967)Kadanoff, G{\"o}tze, Hamblen, Hecht, Lewis,
  Palciauskas et~al.}]{kadanoff67}
Kadanoff LP, G{\"o}tze W, Hamblen D, Hecht R, Lewis E, Palciauskas VV, et~al.
\newblock Static phenomena near critical points: theory and experiment.
\newblock {\em Reviews of Modern Physics\/} {\bf 39} (1967) 395.

\bibitem[{Kadanoff(2000)}]{kadanoff00}
Kadanoff LP.
\newblock {\em Statistical physics: statics, dynamics and renormalization\/}
  (Singapore: World Scientific Publishing Company) (2000).

\bibitem[{Kenna et~al.(2006)Kenna, Johnston, and Janke}]{Kenna06}
Kenna R, Johnston DA, Janke W.
\newblock Self-consistent scaling theory for logarithmic-correction exponents.
\newblock {\em Phys. Rev. Lett.\/} {\bf 97} (2006) 155702.
\newblock \doi{10.1103/PhysRevLett.97.155702}.

\bibitem[{Kenna and Ruiz-Lorenzo(2008)}]{Kenna08}
Kenna R, Ruiz-Lorenzo JJ.
\newblock Scaling analysis of the site-diluted ising model in two dimensions.
\newblock {\em Phys. Rev. E\/} {\bf 78} (2008) 031134.
\newblock \doi{10.1103/PhysRevE.78.031134}.

\bibitem[{Rub\'{\i} et~al.(2004)Rub\'{\i}, {Santamar\'{\i}a-Holek}, and
  {P\'erez-Madrid}}]{Rubi04}
Rub\'{\i} JM, {Santamar\'{\i}a-Holek} I, {P\'erez-Madrid} A.
\newblock Slow dynamics and local quasi-equilibrium - relaxation in supercooled
  colloidal systems.
\newblock {\em J. Phys.: Condens. Matter\/} {\bf 16} (2004) S2047.
\newblock \doi{10.1088/0953-8984/16/22/002}.

\bibitem[{{Santamar\'{\i}a-Holek} et~al.(2004){Santamar\'{\i}a-Holek},
  {P\'erez-Madrid}, and Rub\'{\i}}]{Santamaria-Holek04}
{Santamar\'{\i}a-Holek} I, {P\'erez-Madrid} A, Rub\'{\i} JM.
\newblock Local quasi-equilibrium description of slow relaxation systems.
\newblock {\em J. Chem. Phys.\/} {\bf 120} (2004) 2818.
\newblock \doi{10.1063/1.1640346}.

\bibitem[{Vainstein et~al.(2003)Vainstein, Stariolo, and
  Arenzon}]{Vainstein03a}
Vainstein MH, Stariolo DA, Arenzon JJ.
\newblock Heterogeneities in systems with quenched disorder.
\newblock {\em J. Phys. A: Math. Gen.\/} {\bf 36} (2003) 10907--10919.
\newblock \doi{10.1088/0305-4470/36/43/016}.

\bibitem[{Santos et~al.(2000)Santos, Oliveira, and Neto}]{Santos00}
Santos MBL, Oliveira EA, Neto AMF.
\newblock Rayleigh scattering of a new lyotropic nematic liquid crystal system:
  crossover of propagative and diffusive behavior.
\newblock {\em Liq. Cryst.\/} {\bf 27} (2000) 1485.

\bibitem[{Benmouna et~al.(2001)Benmouna, Peng, Gapinski, Patkowski, Ruhe, and
  Johannsmann}]{Benmouna01}
Benmouna F, Peng B, Gapinski J, Patkowski A, Ruhe J, Johannsmann D.
\newblock Dynamic light scattering from liquid crystal polymer brushes swollen
  in a nematic solvent.
\newblock {\em Liq. Cryst.\/} {\bf 28} (2001) 1353.

\bibitem[{Peyrard(2001)}]{Peyrard01}
Peyrard M.
\newblock Glass transition in protein hydration water.
\newblock {\em Phys. Rev. E\/} {\bf 64} (2001) 011109.

\bibitem[{Colaiori and Moore(2001{\natexlab{a}})}]{Colaiori01}
Colaiori F, Moore MA.
\newblock {Stretched exponential relaxation in the mode-coupling theory for the
  Kardar-Parisi-Zhang equation}.
\newblock {\em Phys. Rev. E\/} {\bf 63} (2001{\natexlab{a}}) 057103.

\bibitem[{Ferreira et~al.(1991)Ferreira, Ludwig, and Montes}]{Ferreira91}
Ferreira JL, Ludwig GO, Montes A.
\newblock Experimental investigations of ion-acoustic double-layers in the
  electron flow across multidipole magnetic fields.
\newblock {\em Plasma Phys. Control. Fusion\/} {\bf 33} (1991) 297--311.

\bibitem[{Bouchaud et~al.(1991)Bouchaud, M\'ezard, and Yedidia}]{Bouchaud91}
Bouchaud JP, M\'ezard M, Yedidia JS.
\newblock Variational theory for disordered vortex lattices.
\newblock {\em Phys. Rev. Lett.\/} {\bf 67} (1991) 3840.

\bibitem[{Kohlrausch(1854)}]{Kohlrausch54}
Kohlrausch R.
\newblock Theorie des elektrischen {R}ückstandes in der {L}eidener {F}lasche.
\newblock {\em Annalen der Physik\/} {\bf 167} (1854) 56--82.
\newblock \doi{10.1002/andp.18541670103}.

\bibitem[{Kohlrausch(1863)}]{Kohlrausch63}
Kohlrausch F.
\newblock {Über die elastische Nachwirkung bei der Torsion}.
\newblock {\em Annalen der Physik\/} {\bf 195} (1863) 337.
\newblock \doi{10.1002/andp.18631950702}.

\bibitem[{Lapas et~al.(2015)Lapas, Ferreira, Rub{\'\i}, and Oliveira}]{Lapas15}
Lapas LC, Ferreira RM, Rub{\'\i} JM, Oliveira FA.
\newblock Anomalous law of cooling.
\newblock {\em J. Chem. Phys.\/} {\bf 142} (2015) 104106.
\newblock \doi{10.1063/1.4914872}.

\bibitem[{Mittag-Leffler(1905)}]{Mittag-Leffler05}
Mittag-Leffler GM.
\newblock Sur la représentation analytique d'une branche uniforme d'une
  fonction monogène.
\newblock {\em Acta Math.\/} {\bf 29} (1905) 101.

\bibitem[{Khinchin(1949)}]{Khinchin49}
Khinchin AI.
\newblock {\em {M}athematical {F}oundations of {S}tatistical {M}echanics\/}
  (New York: Dover) (1949).

\bibitem[{Bao et~al.(2005)Bao, H\"anggi, and Zhuo}]{Bao05a}
Bao JD, H\"anggi P, Zhuo YZ.
\newblock Non-markovian {B}rownian dynamics and nonergodicity.
\newblock {\em Phys. Rev. E\/} {\bf 72} (2005) 061107.
\newblock \doi{10.1103/PhysRevE.72.061107}.

\bibitem[{Silvestre and Rocha~Filho(2016)}]{Silvestre}
Silvestre C, Rocha~Filho T.
\newblock Ergodicity in a two-dimensional self-gravitating many-body system.
\newblock {\em Phys. Lett. A\/} {\bf 380} (2016) 337--348.

\bibitem[{Campa et~al.(2009)Campa, Dauxois, and Ruffo}]{Campa09}
Campa A, Dauxois T, Ruffo S.
\newblock Statistical mechanics and dynamics of solvable models with long-range
  interactions.
\newblock {\em Phys. Rep.\/} {\bf 480} (2009) 57--159.

\bibitem[{Parisi(1997)}]{Parisi97}
Parisi G.
\newblock {Off-Equilibrium Fluctuation-Dissipation Relation in Fragile
  Glasses}.
\newblock {\em Phys. Rev. Lett.\/} {\bf 79} (1997) 3660--3663.
\newblock \doi{10.1103/PhysRevLett.79.3660}.

\bibitem[{Dybiec et~al.(2012)Dybiec, Parrondo, and Gudowska-Nowak}]{Dybiec12}
Dybiec B, Parrondo JMR, Gudowska-Nowak E.
\newblock Fluctuation-dissipation relations under {L}évy noises.
\newblock {\em EPL\/} {\bf 98} (2012) 50006.

\bibitem[{Cugliandolo et~al.(1994)Cugliandolo, Kurchan, and
  Parisi}]{cugliandolo94}
Cugliandolo L, Kurchan J, Parisi G.
\newblock Off equilibrium dynamics and aging in unfrustrated systems.
\newblock {\em Journal de Physique I\/} {\bf 4} (1994) 1641--1656.

\bibitem[{Oldham and Spanier(1974)}]{oldham74}
Oldham K, Spanier J.
\newblock {\em The fractional calculus theory and applications of
  differentiation and integration to arbitrary order\/}, vol. 111 (San Diego:
  Elsevier) (1974).

\bibitem[{Kilbas et~al.(2006)Kilbas, Srivastava, and Trujillo}]{kilbas06}
Kilbas AAA, Srivastava HM, Trujillo JJ.
\newblock {\em Theory and applications of fractional differential equations\/},
  vol. 204 (Amsterdam: Elsevier Science Limited) (2006).

\bibitem[{Klafter et~al.(1996)Klafter, Shlesinger, and Zumofen}]{Klafter96}
Klafter J, Shlesinger MF, Zumofen G.
\newblock Beyond {B}rownian motion.
\newblock {\em Phys. Today\/} {\bf 49} (1996) 33.

\bibitem[{Edwards and Wilkinson(1982)}]{Edwards82}
Edwards SF, Wilkinson DR.
\newblock The surface statistics of a granular aggregate.
\newblock {\em Proc. R. Soc. Lond. A\/} {\bf 381} (1982) 17--31.
\newblock \doi{10.1098/rspa.1982.0056}.

\bibitem[{Kardar et~al.(1986)Kardar, Parisi, and Zhang}]{Kardar86}
Kardar M, Parisi G, Zhang YC.
\newblock Dynamic scaling of growing interfaces.
\newblock {\em Phys. Rev. Lett.\/} {\bf 56} (1986) 889--892.
\newblock \doi{10.1103/PhysRevLett.56.889}.

\bibitem[{Hansen et~al.(2000)Hansen, Schmittbuhl, Batrouni, and
  de~Oliveira}]{Hansen00}
Hansen A, Schmittbuhl J, Batrouni GG, de~Oliveira FA.
\newblock Normal stress distribution of rough surfaces in contact.
\newblock {\em Geophys. Res. Lett.\/} {\bf 27} (2000) 3639--3642.

\bibitem[{Cordeiro et~al.(2001)Cordeiro, Lima, Dias, and Oliveira}]{Cordeiro01}
Cordeiro JA, Lima MVBT, Dias RM, Oliveira FA.
\newblock Morphology of growth by random walk deposition.
\newblock {\em Physica A\/} {\bf 295} (2001) 209.

\bibitem[{Schmittbuhl et~al.(2006)Schmittbuhl, Chambon, Hansen, and
  Bouchon}]{Schmittbuhl06}
Schmittbuhl J, Chambon G, Hansen A, Bouchon M.
\newblock Are stress distributions along faults the signature of asperity
  squeeze?
\newblock {\em Geophys. Res. Lett.\/} {\bf 33} (2006).

\bibitem[{Horowitz et~al.(2001)Horowitz, Monetti, and Albano}]{Horowitz01}
Horowitz CM, Monetti RA, Albano EV.
\newblock Competitive growth model involving random deposition and random
  deposition with surface relaxation.
\newblock {\em Phys. Rev. E\/} {\bf 63} (2001) 066132.
\newblock \doi{10.1103/PhysRevE.63.066132}.

\bibitem[{Henkel and Durang(2015)}]{Henkel15}
Henkel M, Durang X.
\newblock Spherical model of growing interfaces.
\newblock {\em J. Stat. Mech. Theory Exp.\/} {\bf 2015} (2015) P05022.
\newblock \doi{10.1088/1742-5468/2015/05/p05022}.

\bibitem[{Hairer(2013)}]{Hairer13}
Hairer M.
\newblock Solving the {KPZ} equation.
\newblock {\em Ann. Math.\/} {\bf 178} (2013) 559--664.
\newblock \doi{10.4007/annals.2013.178.2.4}.

\bibitem[{Sasamoto and Spohn(2010)}]{Sasamoto10}
Sasamoto T, Spohn H.
\newblock One-dimensional {Kardar-Parisi-Zhang} equation: An exact solution and
  its universality.
\newblock {\em Phys. Rev. Lett.\/} {\bf 104} (2010) 230602.
\newblock \doi{10.1103/PhysRevLett.104.230602}.

\bibitem[{Kardar(1985)}]{Kardar85}
Kardar M.
\newblock Roughening by impurities at finite temperatures.
\newblock {\em Phys. Rev. Lett.\/} {\bf 55} (1985) 2923.
\newblock \doi{10.1103/PhysRevLett.55.2923}.

\bibitem[{Bertini and Giacomin(1997)}]{Bertine97}
Bertini L, Giacomin G.
\newblock Stochastic burgers and {KPZ} equations from particle systems.
\newblock {\em Commun. Math. Phys.\/} {\bf 183} (1997) 571--607.

\bibitem[{Spitzer(1970)}]{Spitzer70}
Spitzer F.
\newblock Interaction of {M}arkov processes.
\newblock {\em Adv. Math.\/} {\bf 5} (1970) 246 -- 290.
\newblock \doi{10.1016/0001-8708(70)90034-4}.

\bibitem[{\'Odor et~al.(2010)\'Odor, Liedke, and Heinig}]{Odor10}
\'Odor G, Liedke B, Heinig KH.
\newblock Directed $d$-mer diffusion describing the {Kardar-Parisi-Zhang}-type
  surface growth.
\newblock {\em Phys. Rev. E\/} {\bf 81} (2010) 031112.
\newblock \doi{10.1103/PhysRevE.81.031112}.

\bibitem[{Myllys et~al.(2001)Myllys, Maunuksela, Alava, Ala-Nissila, Merikoski,
  and Timonen}]{Mylles01}
Myllys M, Maunuksela J, Alava M, Ala-Nissila T, Merikoski J, Timonen J.
\newblock Kinetic roughening in slow combustion of paper.
\newblock {\em Phys. Rev. E\/} {\bf 64} (2001) 036101.

\bibitem[{Myllys et~al.(2003)Myllys, Maunuksela, Merikoski, Timonen, Horvath,
  Ha et~al.}]{Mylles03}
Myllys M, Maunuksela J, Merikoski J, Timonen J, Horvath V, Ha M, et~al.
\newblock Effect of a columnar defect on the shape of slow-combustion fronts.
\newblock {\em Phys. Rev. E\/} {\bf 68} (2003) 051103.

\bibitem[{Merikoski et~al.(2003)Merikoski, Maunuksela, Myllys, Timonen, and
  Alava}]{Merikoski03}
Merikoski J, Maunuksela J, Myllys M, Timonen J, Alava MJ.
\newblock Temporal and spatial persistence of combustion fronts in paper.
\newblock {\em Phys. Rev. Lett.\/} {\bf 90} (2003) 24501.

\bibitem[{Csah\'ok and Vicsek(1992)}]{Csahok92}
Csah\'ok Z, Vicsek T.
\newblock Kinetic roughening in a model of sedimentation of granular materials.
\newblock {\em Phys. Rev. A\/} {\bf 46} (1992) 4577--4581.
\newblock \doi{10.1103/PhysRevA.46.4577}.

\bibitem[{Ben-Jacob et~al.(1994)Ben-Jacob, Schochet, Tenenbaum, Cohen, Czirók,
  and Vicsek}]{Ben-Jacob94}
Ben-Jacob E, Schochet O, Tenenbaum A, Cohen I, Czirók A, Vicsek T.
\newblock Generic modelling of cooperative growth patterns in bacterial
  colonies.
\newblock {\em Nature\/} {\bf 368} (1994) 46--49.
\newblock \doi{10.1038/368046a0}.

\bibitem[{Matsushita and Fujikawa(1990)}]{Matsushita90}
Matsushita M, Fujikawa H.
\newblock Diffusion-limited growth in bacterial colony formation.
\newblock {\em Physica A\/} {\bf 168} (1990) 498 -- 506.
\newblock \doi{10.1016/0378-4371(90)90402-E}.

\bibitem[{Takeuchi et~al.(2011)Takeuchi, Sano, Sasamoto, and
  Spohn}]{Takeuchi11}
Takeuchi KA, Sano M, Sasamoto T, Spohn H.
\newblock Growing interfaces uncover universal fluctuations behind scale
  invariance.
\newblock {\em Sci. Rep.\/} {\bf 1} (2011) 34.

\bibitem[{Takeuchi and Sano(2012)}]{Takeuchi12}
Takeuchi KA, Sano M.
\newblock Evidence for geometry-dependent universal fluctuations of the
  {Kardar-Parisi-Zhang} interfaces in liquid-crystal turbulence.
\newblock {\em J. Stat. Phys.\/} {\bf 147} (2012) 853--890.

\bibitem[{Takeuchi(2013)}]{Takeuchi13}
Takeuchi KA.
\newblock Crossover from growing to stationary interfaces in the
  {Kardar-Parisi-Zhang} class.
\newblock {\em Phys. Rev. Lett.\/} {\bf 110} (2013) 210604.

\bibitem[{Almeida et~al.(2017)Almeida, Ferreira, Ferraz, and
  Oliveira}]{Almeida17}
Almeida RAL, Ferreira SO, Ferraz I, Oliveira TJ.
\newblock Initial pseudo-steady state \& asymptotic kpz universality in
  semiconductor on polymer deposition.
\newblock {\em Sci. Rep.\/} {\bf 7} (2017) 3773.
\newblock \doi{10.1038/s41598-017-03843-1}.

\bibitem[{Mello et~al.(2001)Mello, Chaves, and Oliveira}]{Mello01}
Mello BA, Chaves AS, Oliveira FA.
\newblock Discrete atomistic model to simulate etching of a crystalline solid.
\newblock {\em Phys. Rev. E\/} {\bf 63} (2001) 041113.
\newblock \doi{10.1103/PhysRevE.63.041113}.

\bibitem[{{Aar\~ao Reis}(2003)}]{Reis03}
{Aar\~ao Reis} F.
\newblock Dynamic transition in etching with poisoning.
\newblock {\em Phys. Rev. E.\/} {\bf 68} (2003) 041602.

\bibitem[{{Aar\~ao Reis}(2004)}]{Reis04}
{Aar\~ao Reis} F.
\newblock Universality in two-dimensional {Kardar-Parisi-Zhang} growth.
\newblock {\em Phys. Rev. E.\/} {\bf 69} (2004) 021610.

\bibitem[{{Aar\~ao Reis}(2005)}]{Reis05}
{Aar\~ao Reis} FDA.
\newblock Numerical study of roughness distributions in nonlinear models of
  interface growth.
\newblock {\em Phys. Rev. E\/} {\bf 72} (2005) 032601.
\newblock \doi{10.1103/PhysRevE.72.032601}.

\bibitem[{Oliveira and Aar\~ao Reis(2008)}]{Oliveira08}
Oliveira TJ, Aar\~ao Reis FDA.
\newblock Maximal- and minimal-height distributions of fluctuating interfaces.
\newblock {\em Phys. Rev. E\/} {\bf 77} (2008) 041605.
\newblock \doi{10.1103/PhysRevE.77.041605}.

\bibitem[{Tang et~al.(2010)Tang, Xun, Wen, Han, Xia, Hao et~al.}]{Tang10}
Tang G, Xun Z, Wen R, Han K, Xia H, Hao D, et~al.
\newblock Discrete growth models on deterministic fractal substrate.
\newblock {\em Physica A\/} {\bf 389} (2010) 4552--4557.

\bibitem[{Xun et~al.(2012)Xun, Zhang, Li, Xia, Hao, and Tang}]{Xun12}
Xun Z, Zhang Y, Li Y, Xia H, Hao D, Tang G.
\newblock Dynamic scaling behaviors of the discrete growth models on fractal
  substrates.
\newblock {\em J. Stat. Mech. Theory Exp.\/} {\bf 2012} (2012) P10014.

\bibitem[{Rodrigues et~al.(2015)Rodrigues, Mello, and Oliveira}]{Rodrigues15}
Rodrigues EA, Mello BA, Oliveira FA.
\newblock Growth exponents of the etching model in high dimensions.
\newblock {\em J. Phys. A\/} {\bf 48} (2015) 35001--35012.
\newblock \doi{10.1088/1751-8113/48/3/035001}.

\bibitem[{{Mello}(2015)}]{Mello15}
{Mello} BA.
\newblock {A random rule model of surface growth}.
\newblock {\em Physica A\/} {\bf 419} (2015) 762--767.
\newblock \doi{10.1016/j.physa.2014.10.064}.

\bibitem[{Alves et~al.(2016)Alves, Rodrigues, Fernandes, Mello, Oliveira, and
  Costa}]{Alves16}
Alves WS, Rodrigues EA, Fernandes HA, Mello BA, Oliveira FA, Costa IVL.
\newblock Analysis of etching at a solid-solid interface.
\newblock {\em Phys. Rev. E\/} {\bf 94} (2016) 042119.

\bibitem[{Carrasco and Oliveira(2016)}]{Carrasco16}
Carrasco ISS, Oliveira TJ.
\newblock Universality and dependence on initial conditions in the class of the
  nonlinear molecular beam epitaxy equation.
\newblock {\em Phys. Rev. E\/} {\bf 94} (2016) 050801.
\newblock \doi{10.1103/PhysRevE.94.050801}.

\bibitem[{Carrasco and Oliveira(2018)}]{Carrasco18}
Carrasco ISS, Oliveira TJ.
\newblock {Kardar-Parisi-Zhang} growth on one-dimensional decreasing
  substrates.
\newblock {\em Phys. Rev. E\/} {\bf 98} (2018) 010102.
\newblock \doi{10.1103/PhysRevE.98.010102}.

\bibitem[{Henkel(2017)}]{Henkel17}
Henkel M.
\newblock From dynamical scaling to local scale-invariance: a tutorial.
\newblock {\em The European Physical Journal Special Topics\/} {\bf 226} (2017)
  605--625.
\newblock \doi{10.1140/epjst/e2016-60336-5}.

\bibitem[{Henkel et~al.(2012)Henkel, Noh, and Pleimling}]{Henkel12}
Henkel M, Noh JD, Pleimling M.
\newblock Phenomenology of aging in the kardar-parisi-zhang equation.
\newblock {\em Phys. Rev. E\/} {\bf 85} (2012) 030102.
\newblock \doi{10.1103/PhysRevE.85.030102}.

\bibitem[{Kelling et~al.(2017{\natexlab{a}})Kelling, Ódor, and
  Gemming}]{Kelling17a}
Kelling J, Ódor G, Gemming S.
\newblock Local scale-invariance of the 2+1 dimensional
  {Kardar–Parisi–Zhang model}.
\newblock {\em J. Phys. A\/} {\bf 50} (2017{\natexlab{a}}) 12LT01.

\bibitem[{Wio et~al.(2010)Wio, Revelli, Deza, Escudero, and
  de~La~Lama}]{Wio10b}
Wio HS, Revelli JA, Deza RR, Escudero C, de~La~Lama MS.
\newblock Discretization-related issues in the {Kardar-Parisi-Zhang} equation:
  Consistency, galilean-invariance violation, and fluctuation-dissipation
  relation.
\newblock {\em Phys. Rev. E\/} {\bf 81} (2010) 066706.
\newblock \doi{10.1103/PhysRevE.81.066706}.

\bibitem[{Wio et~al.(2017)Wio, Rodr{\'\i}guez, Gallego, Revelli, Al{\'e}s, and
  Deza}]{Wio17}
Wio HS, Rodr{\'\i}guez MA, Gallego R, Revelli JA, Al{\'e}s A, Deza RR.
\newblock d-dimensional {KPZ} equation as a stochastic gradient flow in an
  evolving landscape: Interpretation and time evolution of its generating
  functional.
\newblock {\em Frontiers in Physics\/} {\bf 4} (2017) 52.

\bibitem[{Colaiori and Moore(2001{\natexlab{b}})}]{Francesca01}
Colaiori F, Moore M.
\newblock Upper critical dimension, dynamic exponent, and scaling functions in
  the mode-coupling theory for the {Kardar-Parisi-Zhang} equation.
\newblock {\em Phys. Rev. Lett.\/} {\bf 86} (2001{\natexlab{b}}) 3946.

\bibitem[{Schwartz and Perlsman(2012)}]{Schwartz12}
Schwartz M, Perlsman E.
\newblock Upper critical dimension of the {Kardar-Parisi-Zhang} equation.
\newblock {\em Phys. Rev. E\/} {\bf 85} (2012) 050103.

\bibitem[{Lam and Shin(1998)}]{lam98}
Lam CH, Shin FG.
\newblock Improved discretization of the {Kardar-Parisi-Zhang} equation.
\newblock {\em Phys. Rev. E\/} {\bf 58} (1998) 5592.

\bibitem[{Xu et~al.(2006)Xu, Han, and Wu}]{Xu06}
Xu Z, Han H, Wu X.
\newblock Numerical method for the deterministic {Kardar-Parisi-Zhang} equation
  in unbounded domains.
\newblock {\em Commun. Comput. Phys\/} {\bf 1} (2006) 479--493.

\bibitem[{Halpin-Healy and Takeuchi(2015)}]{HalpinHealy15}
Halpin-Healy T, Takeuchi KA.
\newblock A {KPZ} cocktail-shaken, not stirred...
\newblock {\em J. Stat. Phys.\/} {\bf 160} (2015) 794--814.

\bibitem[{Torres and Buceta(2018)}]{Torres17}
Torres MF, Buceta RC.
\newblock Numerical integration of {KPZ} equation with restrictions.
\newblock {\em J.Stat. Mech. Theory Exp.\/} {\bf 2018} (2018) 033208.

\bibitem[{Kelling et~al.(2017{\natexlab{b}})Kelling, {\'O}dor, and
  Gemming}]{Kelling17}
Kelling J, {\'O}dor G, Gemming S.
\newblock Dynamical universality classes of simple growth and lattice gas
  models.
\newblock {\em J. Phys A\/} {\bf 51} (2017{\natexlab{b}}) 035003.
\newblock \doi{10.1088/1751-8121/aa97f3}.

\bibitem[{P\ifmmode~\check{r}\else \v{r}\fi{}edota and
  Kotrla(1996)}]{Predota96}
P\ifmmode~\check{r}\else \v{r}\fi{}edota M, Kotrla M.
\newblock Stochastic equations for simple discrete models of epitaxial growth.
\newblock {\em Phys. Rev. E\/} {\bf 54} (1996) 3933--3942.
\newblock \doi{10.1103/PhysRevE.54.3933}.

\bibitem[{Chua et~al.(2005)Chua, Haselwandter, Baggio, and Vvedensky}]{Chua05}
Chua ALS, Haselwandter CA, Baggio C, Vvedensky DD.
\newblock Langevin equations for fluctuating surfaces.
\newblock {\em Phys. Rev. E\/} {\bf 72} (2005) 051103.

\bibitem[{Buceta et~al.(2014)Buceta, Hansmann, and {von Haeften}}]{Buceta14}
Buceta RC, Hansmann D, {von Haeften} B.
\newblock {Revisiting random deposition with surface relaxation: approaches
  from growth rules to the Edwards-Wilkinson equation}.
\newblock {\em J. Stat. Mech. Theory Exp.\/} {\bf 2014} (2014) p12028.

\bibitem[{Pr\"ahofer and Spohn(2000)}]{Prahofer00}
Pr\"ahofer M, Spohn H.
\newblock Universal distributions for growth processes in $1+1$ dimensions and
  random matrices.
\newblock {\em Phys. Rev. Lett.\/} {\bf 84} (2000) 4882--4885.
\newblock \doi{10.1103/PhysRevLett.84.4882}.

\bibitem[{Johansson(2000)}]{Johansson00}
Johansson K.
\newblock Shape fluctuations and random matrices.
\newblock {\em Commun. Math. Phys.\/} {\bf 209} (2000) 437--476.

\bibitem[{Oliveira et~al.(2013)Oliveira, Alves, and Ferreira}]{Oliveira13}
Oliveira TJ, Alves SG, Ferreira SC.
\newblock {Kardar-Parisi-Zhang }universality class in (2+1) dimensions:
  Universal geometry-dependent distributions and finite-time corrections.
\newblock {\em Phys. Rev. E\/} {\bf 87} (2013) 040102.

\bibitem[{Alves et~al.(2013)Alves, Oliveira, and Ferreira}]{Alves13}
Alves SG, Oliveira TJ, Ferreira SC.
\newblock Non-universal parameters, corrections and universality in
  {Kardar--Parisi--Zhang} growth.
\newblock {\em J. Stat. Mech. Theory Exp.\/} {\bf 2013} (2013) P05007.

\bibitem[{Almeida et~al.(2014)Almeida, Ferreira, Oliveira, and {Aar\~ao
  Reis}}]{Almeida14}
Almeida RAL, Ferreira SO, Oliveira TJ, {Aar\~ao Reis} FDA.
\newblock Universal fluctuations in the growth of semiconductor thin films.
\newblock {\em Phys. Rev. B\/} {\bf 89} (2014) 045309.

\bibitem[{Halpin-Healy and Zhang(1995)}]{Halpin-Healy95}
Halpin-Healy T, Zhang YC.
\newblock Kinetic roughening phenomena, stochastic growth, directed polymers
  and all that. aspects of multidisciplinary statistical mechanics.
\newblock {\em Phys. Rep.\/} {\bf 254} (1995) 215--414.

\bibitem[{Haselwandter and Vvedensky(2006)}]{Haselwandter06}
Haselwandter CA, Vvedensky DD.
\newblock Scaling of ballistic deposition from a {L}angevin equation.
\newblock {\em Phys. Rev. E\/} {\bf 73} (2006) 040101.

\bibitem[{Haselwandter and Vvedensky(2008)}]{Haselwandter08}
Haselwandter CA, Vvedensky DD.
\newblock Renormalization of stochastic lattice models: Epitaxial surfaces.
\newblock {\em Phys. Rev. E\/} {\bf 77} (2008) 061129.

\bibitem[{Silveira and {Aar\~ao Reis}(2012)}]{Silveira12}
Silveira FA, {Aar\~ao Reis} FDA.
\newblock Langevin equations for competitive growth models.
\newblock {\em Phys. Rev. E\/} {\bf 85} (2012) 011601.

\bibitem[{Alcaraz and Bariev(1999)}]{Alcaraz99}
Alcaraz FC, Bariev RZ.
\newblock Exact solution of the asymmetric exclusion model with particles of
  arbitrary size.
\newblock {\em Phys. Rev. E\/} {\bf 60} (1999) 79.

\bibitem[{Meakin(1993)}]{Meakin93}
Meakin P.
\newblock The growth of rough surfaces and interfaces.
\newblock {\em Phys. Rep.\/} {\bf 235} (1993) 189--289.

\bibitem[{Krug(1997)}]{Krug97}
Krug J.
\newblock Origins of scale invariance in growth processes.
\newblock {\em Adv. Phys.\/} {\bf 46} (1997) 139--282.

\bibitem[{Ben-Avraham and Havlin(2000)}]{ben00}
Ben-Avraham D, Havlin S.
\newblock {\em Diffusion and reactions in fractals and disordered systems\/}
  (Cambridge: Cambridge University Press) (2000).

\bibitem[{Abad et~al.(2002)Abad, Masser, and Ben-Avraham}]{abad02}
Abad E, Masser T, Ben-Avraham D.
\newblock Lattice kinetics of diffusion-limited coalescence and annihilation
  with sources.
\newblock {\em J. Phys. A Math. Gen.\/} {\bf 35} (2002) 1483.

\bibitem[{Shapoval et~al.(2018)Shapoval, Dudka, Durang, and
  Henkel}]{shapoval18}
Shapoval D, Dudka M, Durang X, Henkel M.
\newblock Cross-over between diffusion-limited and reaction-limited regimes in
  the coagulation-diffusion process.
\newblock {\em J. Phys. A Math. Theor.\/} {\bf 51} (2018) 425002,
  arXiv:1801.09216.

\bibitem[{Doering and Ben-Avraham(1989)}]{doering89}
Doering CR, Ben-Avraham D.
\newblock Diffusion-limited coagulation in the presence of particle input:
  exact results in one dimension.
\newblock {\em Phys. Rev. Lett.\/} {\bf 62} (1989) 2563.

\bibitem[{Krebs et~al.(1995)Krebs, Pfannm{\"u}ller, Wehefritz, and
  Hinrichsen}]{krebs95}
Krebs K, Pfannm{\"u}ller MP, Wehefritz B, Hinrichsen H.
\newblock {Finite-size scaling studies of one-dimensional reaction-diffusion
  systems. Part I. Analytical results}.
\newblock {\em J. Stat. Phys.\/} {\bf 78} (1995) 1429--1470.

\bibitem[{Simon(1995)}]{simon95}
Simon H.
\newblock Concentration for one and two-species one-dimensional
  reaction-diffusion systems.
\newblock {\em J. Phys. A Math. Gen.\/} {\bf 28} (1995) 6585.

\bibitem[{Evans and Majumdar(2011)}]{evans11}
Evans MR, Majumdar SN.
\newblock Diffusion with stochastic resetting.
\newblock {\em Phys. Rev. Lett.\/} {\bf 106} (2011) 160601.

\bibitem[{Evans and Majumdar(2014)}]{evans14}
Evans MR, Majumdar SN.
\newblock Diffusion with resetting in arbitrary spatial dimension.
\newblock {\em J. Phys. A Math. Theor.\/} {\bf 47} (2014) 285001,
  arXiv:1404.4574.

\bibitem[{Durang et~al.(2014)Durang, Henkel, and Park}]{durang14}
Durang X, Henkel M, Park H.
\newblock The statistical mechanics of the coagulation--diffusion process with
  a stochastic reset.
\newblock {\em J. Phys. A Math. Theor.\/} {\bf 47} (2014) 045002,
  arxiv:1309.2107.

\bibitem[{Hodge(1995)}]{hodge95}
Hodge IM.
\newblock Physical aging in polymer glasses.
\newblock {\em Science\/} {\bf 267} (1995) 1945--1947.

\end{thebibliography}

\end{document}